\documentclass[]{aastex631}
\usepackage{amsmath}
\usepackage{amssymb}
\usepackage{hyperref}
\usepackage{parskip}
\usepackage{verbatim}
\usepackage{graphicx}
\usepackage{amssymb}
\begin{document}

\title{X-ray spectral and temporal evolution of atoll source 4U 1820-30 with \textit{AstroSat}: detection of high frequency quasi-periodic oscillation}

\author[0000-0001-6877-4393]{SUBHASISH DAS}
\affiliation{Department of Pure and Applied Physics, Guru Ghasidas Vishwavidyalaya (A Central University), Bilaspur (C. G.)—495009, India.}

\author{Vivek K. Agrawal}
\affiliation{Space Astronomy Group, ISITE Campus, U. R. Rao Satellite Center, Outer Ring Road, Marathahalli, Bangalore—560037, India}

\correspondingauthor{Parijat Thakur}
\email{parijat@associates.iucaa.in, pthakurncu@gmail.com}
\author[0000-0001-5958-0562]{Parijat Thakur}
\affiliation{Department of Pure and Applied Physics, Guru Ghasidas Vishwavidyalaya (A Central University), Bilaspur (C. G.)—495009, India.}

\author[0000-0003-1589-2075]{G. C. Dewangan}
\affiliation{Inter-University Centre for Astronomy and Astrophysics, Post Bag 4, Ganeshkhind, Pune—411007, India}

\author[0000-0003-3469-9895]{Raj Kumar}
\affiliation{Department of Astronomy, Xiamen University, Xiamen, Fujian 361005, People's Republic of China}

\author[0009-0005-4510-4051]{Pragati Sahu}
\affiliation{Department of Pure and Applied Physics, Guru Ghasidas Vishwavidyalaya (A Central University), Bilaspur (C. G.)—495009, India.}

\author[0000-0002-0638-950X]{Vineet Kumar Mannaday}
\affiliation{Department of Physics, Govt. Niranjan Kesharwani College, Kota, Bilaspur
(C.G.)-495113, India}





\begin{abstract}
AstroSat/LAXPC and SXT observed the persistent neutron star low-mass X-ray binary 4U 1820-30 between 2016 and 2022. During these observations, the hardness-intensity diagram (HID) and color-color diagram (CCD) indicated that the source was in the banana state. We divided the CCD into 11 segments for spectral and timing analyses. For each segment in the CCD, we modeled the spectral data using two distinct approaches over the 0.7-20.0 keV band. A combination of a multi-color-disk component with an inner disk temperature of around 0.6 keV and Comptonized emission from the boundary layer (BL)/ hot corona provided the best description of the X-ray spectral data of this source. The truncation radius was found to be in the range of $\sim$ 19-40 km. The Comptonized component has an optical depth in the range of $\sim 7 - 13$ with electron temperature in the range of $\sim 2.5 - 3.8$ keV. The optical depth of the corona varies significantly along the position on the CCD, while $\sim$ 80\% of the X-ray flux comes from the Comptonized component. We discuss possible physical scenarios to explain the relationship between the spectral evolution and motion of the source along the CCD. The timing analysis revealed kHz QPOs peaks at $\sim 710$ Hz and $\sim 740$ Hz in the lower left banana branch. An energy-dependent study indicates that these QPOs are stronger in the high-energy band.
\end{abstract}



\section{Introduction} \label{sec:intro}
A neutron star (NS) accretes matter from a low-mass ($\leq 1 M_{\odot}$) companion star via Roche-lobe overflow in NS low-mass X-ray binaries (LMXRBs), and the accreted material forms an accretion disk \citep{Bahramian2022}. NS-LMXRBs host a weakly magnetized ($B\leq 10^{8-9}$G) NS and provide ideal laboratories to understand the physical behavior of accretion processes in the vicinity of a dense compact object and in the strong gravity regime. NS-LMXRBs are categorized into two classes: Z-type and atoll-type sources, based on their X-ray spectral and temporal characteristics as well as the pattern they trace in their color-color diagram (CCD) or hardness-intensity diagram (HID) \citep{Hasinger, DiSalvo:2023aoc}. 
Z-type and atoll-type sources differ in luminosity due to their varied accretion rates \citep{lin2009spectral, homan2010xte, ng2024x}. Z-type sources are the most luminous NS-LMXRBs, and sometimes they radiate at a rate close to the Eddington luminosity ($L_\mathrm{Edd}$), whereas atoll-type sources are generally less bright ($\sim 0.001-0.5\; L_\mathrm{Edd}$) than Z-type sources \citep{done}. Based on luminosity, atoll-type sources are further classified into two categories: ordinary atoll sources (0.01-0.3 $L_\mathrm{Edd}$) and bright atoll sources (0.3-0.5 $L_\mathrm{Edd}$) \citep{Yao_2021}. The spectral states of atoll sources are broadly categorized into: hard state, termed as the `extreme island state' (EIS), the intermediate state, referred to as the `island state' (IS), and the soft state, designated as the `banana state' (BS) \citep{van}. BS is further subdivided into the lower banana (LB), lower-left banana (LLB), and upper banana (UB) branches \citep{wang}. Atoll-type sources exhibit a ``\textbf{C}-shaped" track in the CCD/HID and move along this track on timescales of days to months \citep{Hasinger, van, Fridriksson2015, wang}. Along this track, the mass accretion rate ($\dot m$) increases as the source moves from the IS to the UB state \citep{Falagna2006}.

In the IS state, sources exhibit lower luminosities with harder spectra, while in the BS they show higher luminosities with softer spectra \citep{barret2001broad}. The X-ray spectra of these sources have two major components: a soft thermal component and a hard Comptonized component. To model the X-ray spectra of atoll sources, various approaches have been proposed in the literature. According to one viewpoint, the hard Comptonized component originates from the boundary layer (BL), while the soft thermal emission component is characterized as a multi-color disk (MCD) blackbody that originates from the cold accretion disk \citep{mitsuda1984energy,barret2001broad,di2002spectral,agrawal2003x,agrawal2009x,agrawal20184U170544}.
In an alternative approach, the emission component originating from a hot inner disk is responsible for the hard Comptonized component, and the BL radiates a soft single-temperature blackbody (BB) \citep{di2001study,sleator2016nustar}. In certain cases, the X-ray spectra can be modeled using two thermal components: one from the accretion disk (MCD) and one from the BL (BB) \citep{lin2007evaluating,lin2010suzaku,agrawal20184U170544}.

NS-LMXRBs show fast time variability features in their power density spectrum (PDS) \citep{DiSalvo:2023aoc}.
They exhibit two types of noise features in the PDS: a very low-frequency noise (VLFN) component modeled using a power law function and a band-limited noise (BLN) component modeled using a Lorentzian function. If the BLN centroid frequency is $> 10\;\mathrm{Hz}$, it is called a high-frequency noise (HFN); otherwise, it is called low-frequency noise (LFN) \citep{Jonker1998,Jonker2000,homan2002rxte,Agarwa2020}. Occasionally, narrow peak-like features known as quasi-periodic oscillations (QPOs) are also observed in the PDS \citep{Strohmayer1996, Vanderklis1996, Peirano2021}. QPOs with frequencies ranging from 1 Hz to 70 Hz are named low-frequency QPOs (LF-QPOs) \citep{Homan2015}. However, the high-frequency peak ($\geq$ 200 Hz) is generally termed a kilohertz (kHz) QPO \citep{Van2000, wang}. In most atoll-type sources, kHz QPOs are mostly observed in the LLB branch \citep{wang}. kHz QPOs are often detected in pairs referred to as lower and upper kHz QPOs \citep{van1997kilohertz, Belloni2002, wang}. Their frequencies vary with $\dot m$. At low $\dot m$ in atoll sources, the PDS is dominated by BLN, correlated with the source position on the CCD. In the lowest $\dot m$ regime (IS), the BLN fractional rms is large ($\sim 10-20$\%), falling to $\leq 2\%$ as $\dot m$ increases, and this is the region on the CCD where kHz QPOs appear \citep{wijnands1999broadband,psaltis1999correlations,di2001study, Salvo2003}. At the highest $\dot m$, when the kHz QPOs are no longer observed, VLFN dominates below $\sim 1$ Hz, and BLN components are sometimes detected with a characteristic frequency of 10--20 Hz \citep{Hasinger,di2001study}. 

4U 1820-30 is a persistent ultra-compact (UC) NS-LMXRB \citep{giacconi1974third, Stella, ArmasPadilla2023} consisting of an NS as the primary and a He white dwarf star ($0.06-0.08 M\odot$) as its companion \citep{Rappaport, Smale1997}. It is located at $0.66^{\prime \prime}$ from the center of the NGC 6624  globular cluster \citep{Rappaport, Smale1997, Shaposhnikov2004, Mondal:2016yez, IXPE2023}. \cite{Grindlay} confirmed this source as an NS based on the detection of a type-I thermonuclear burst. \cite{Stella} determined its orbital period to be 11.4 min, the shortest orbital period known for any XRB. This source is classified as a bright atoll-type source \citep{Hasinger,tarana2007integral,iaria2020reflection}, which mainly exhibits BS, with occasional transitions to IS \citep{tarana2007integral}. This source also shows a 170-day super-orbital accretion cycle. 4U 1820-30 also shows an intrinsic luminosity modulation, possibly due to the tidal effects of a third object. As a result, 4U 1820-30 oscillates between low and high luminosity modes, which have distinct spectral features connected with the variation of accretion flow \citep{marino2023accretion, Chou}. \cite{guver2010mass} estimated the mass and radius of the NS at nearly $\sim 1.6 M_\odot$ and 9.11 km, respectively. \cite{anderson1997time} estimated the inclination angle in the range of $35^\circ-50^\circ$. \cite{Baumgardt2021} reported the distance of the source as $D=8.0\pm0.1$ kpc using \textit{GAIA EDR3}. 

The observation of 4U 1820-30 by the Rossi X-ray Timing Explorer (\textit{RXTE}) has revealed the presence of two simultaneous kHz QPOs \citep{Smale1997}. \cite{zhang} observed the highest QPO frequency at $1060\pm20$ Hz, indicating that the disk extends close to the marginally stable orbit. Later, \cite{Titarchuk} observed LF-QPOs, BLN, VLFN, and high-frequency QPOs in this source using \textit{RXTE}. Interestingly, LF-QPO around 6 Hz was occasionally observed in this source during BS, similar to the normal branch oscillation in the Z-type sources \citep{Wijnands, Titarchuk}. Numerous studies have investigated the spectral behavior of 4U 1820-30 with different X-ray instruments \citep{Stella,Smale1994,christian1997survey,piraino1999bepposax,kaaret1999strong}. \cite{Titarchuk} found that the photon index ($\Gamma$) remains constant at $\sim 2$ during the transition from BS to IS. They suggested that the stability of $\Gamma$ is associated with the spectral properties of an accreting NS source. Based on spectral analysis using \textit{NuSTAR}, \textit{Swift}, and \textit{Suzaku} data, \cite{Mondal:2016yez} suggested that seed photons from the geometrically thin accretion disk are Comptonized by the BL region, which covers a large portion of the NS surface. \cite{marino2023accretion} found stability in disk parameters and substantial evolution in the Comptonization component using spectral analysis of \textit{NuSTAR}, \textit{NICER}, and \textit{AstroSat} data. Their results indicated that a varying region that provides the seed photons for the Comptonization spectrum was responsible for the observed X-ray flux modulation. 4U 1820-30 is also known as a radio emitter \citep{Migliari2004,Russell2021,marino2023accretion,IXPE2023}. Recently, \cite{IXPE2023} detected X-ray polarization using \textit{IXPE} and suggested that the polarization may occur in an optically thick outflow emanating from the inner part of the accretion disk.

Spectral and temporal studies of the NS-LMXRB 4U 1820-30 have shown variability over different time scales. 
In this work, we focused on the investigation of the long-term evolution of spectral and temporal properties along the CCD using available \textit{AstroSat} data. The paper is organized as follows. In section  \ref{sec:Observation}, we provide observational details and data reduction techniques. In section \ref{sec:Analysis}, we present the analysis and modeling methods of the temporal and spectral data. In section \ref{Results and discussion}, we present the results. In section \ref{discusion}, we interpret and discuss our results and conclude in the last section.
\begin{center}
\begin{longtable}[H]{lllllllllll}
\caption{Details of \textit{AstroSat} observation for 4U 1820-30 during 2016 to 2022.}
\label{Observation details}\\
\hline
\hline
Epoch & Observation & Observation & Star Date & Stop Date & Total Effective  \\
        &  No.      &  ID    &      &  &Exposure(sec)  \\

	\hline
 \endfirsthead
\multicolumn{9}{c}%
{\tablename\ \thetable\ -- \textit{Continued from previous page}} \\
 \hline
Epoch &Observation & Observation & Star Time & Stop Time & Total Effective  \\
        &  No.      &  ID     &      &  &Exposure(s)  \\

\hline
\endhead
\hline \multicolumn{9}{r}{\textit{Continued on next page}} \\
\endfoot
\hline
\endlastfoot
 Epoch A& Obs 1 & \textit{A02\_098T01\_9000000736} & 2016-10-17 & 2016-10-18 & 93193  \\
 \hline
  & Obs 2& \textit{A03\_107T01\_9000001246} & 2017-05-26 & 2017-05-26 & 15083  \\

& Obs 3 & \textit{A03\_107T01\_9000001418} & 2017-07-30 & 2017-07-30 & 2848  \\
 
& Obs 4  & \textit{G07\_047T01\_9000001424} & 2017-07-31 & 2017-08-02 & 47887 \\
 
 Epoch B& Obs 5 & \textit{A03\_107T01\_9000001530} & 2017-09-09 & 2017-09-10 & 24962  \\

& Obs 6  & \textit{A04\_055T02\_9000001576} & 2017-10-03 & 2017-10-04 & 59458  \\
 
& Obs 7 & \textit{G08\_036T01\_9000001584} & 2017-10-05 & 2017-10-06 & 13173  \\
 
& Obs 8   & \textit{G08\_036T01\_9000002034} & 2018-04-11 & 2018-04-12 & 13216  \\
 
& Obs 9  & \textit{G08\_036T01\_9000002290} & 2018-08-09 & 2018-08-09 & 12820 \\
\hline
Epoch C & Obs 10 & \textit{T05\_022T01\_9000005070} & 2022-04-15 & 2022-04-16 & 47380  \\
 
\end{longtable}
\end{center}

\section{Observational Details and Data Reduction} \label{sec:Observation}
\textit{AstroSat} is the first Indian multi-wavelength satellite \citep{AGRAWAL2006, Singh:2016gau} with four co-aligned instruments that can study astrophysical sources from ultraviolet to hard X-rays. The broadband X-ray energy range of 0.3-80.0 keV is covered by the Large Area X-ray Proportional Counter (LAXPC; \citep{2016yadav, Antia}) and the Soft X-ray Telescope (SXT; \citep{Kpsingh}). \textit{AstroSat} conducted 10 pointed observations (termed Observation ID) of the source 4U 1820-30 between 2016-2022\footnote{\url{https://astrobrowse.issdc.gov.in/astro_archive/archive/Home.jsp}}. We carried out a detailed analysis using the data collected during these observations, whose details are listed in Table \ref{Observation details}. Simultaneous observational data from LAXPC and SXT are used for spectral analysis. We have divided all observations into three parts, Epoch A, Epoch B, and Epoch C, based on the gain change of the \textit{AstroSat}/LAXPC instrument.

\begin{figure}[h!]
    \centering
    \includegraphics[width=0.8\textwidth]{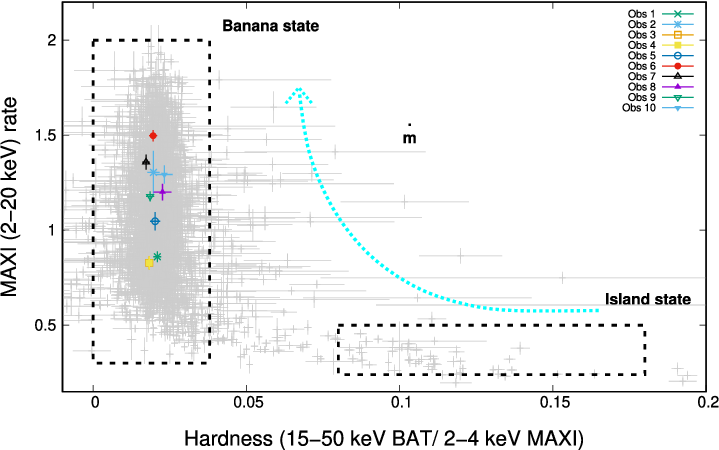}
    \caption{The Hardness Intensity Diagram of 4U 1820-30. We utilized MAXI count rates in the energy range of 2.0–20.0 keV for the intensity. Following \cite{marino2023accretion}, we estimated the hardness as the ratio of the BAT count-rate (15.0–50.0 keV) to the MAXI count-rate in the soft band (2.0–4.0 keV) obtained on the same day. The position in the HID during the dates of the individual AstroSat observations is highlighted with color points superimposed over the grey data points. The positions are labeled with progressive numbers, as shown in Table \ref{Observation details}. ``LB" is at the bottom of the BS, and ``UB" is at the top.}
    \label{maxi_bat_hardness}
\end{figure}

\subsection{SXT}
SXT is capable of observing the X-ray sources in the energy range of 0.3-8.0 keV. The effective area of SXT is 90 $cm^2$ at 1.5 keV, and it has a focal length of 2 meters. SXT has observed the source in photon counting (PC) mode and fast window (FW) mode. The readout time is 2.37 seconds and 278 ms for PC mode and FW mode, respectively. We have considered level-2 data. Using \texttt{SXTMerger.jl\footnote{\url{https://www.tifr.res.in/astrosat_sxt/dataanalysis.html}}} package, we produced an exposure-corrected merged clean event file. Utilizing this event file, we first extracted the source image. In the PC mode, if the source count rate is greater than 40 counts/sec, it indicates pile-up. To avoid this pile-up, we extracted a circular region of radius 12 arcmin, and a region of radius 4 arcmin was excluded from the center of the image. For FW mode, we considered a circular region of 12 arcmin to extract the spectra and light curve. We use background file, and redistribution matrix files (RMF) provided by the SXT team at the Tata Institute of Fundamental Research (TIFR) \footnote{\url{https://www.tifr.res.in/astrosat_sxt/dataanalysis.html}}. To generate vignetting corrected ancillary (ARF) files compatible with the source region, an SXT ARF generation tool \texttt{sxtARFModule\footnote{\url{https://www.tifr.res.in/astrosat_sxt/dataanalysis.html}}} was used. 

\subsection{LAXPC}
The LAXPC instrument consists of three proportional counter detector units named  LAXPC10, LAXPC20, and LAXPC30 with an effective area of 6000 $cm^2$, which can detect X-rays in the energy range of 3-80 keV.
We have used the data only from LAXPC20 because  LAXPC10 and LAXPC30 showed anomalous behavior, including low gain and gas leakage. Using the \texttt{laxpcsoftv3.4.3\_07May2022 \footnote{ \url{https://www.tifr.res.in/~astrosat_laxpc/software.html}}}, we reduced the Level 1 data to generate Level 2 products. Subsequently, source light curves, background light curves and energy spectra were extracted using standard routines available in this software.

\section{ANALYSIS} \label{sec:Analysis}

\subsection{Light curve and CCD}

\begin{figure} [h]
    \centering
\gridline{\fig{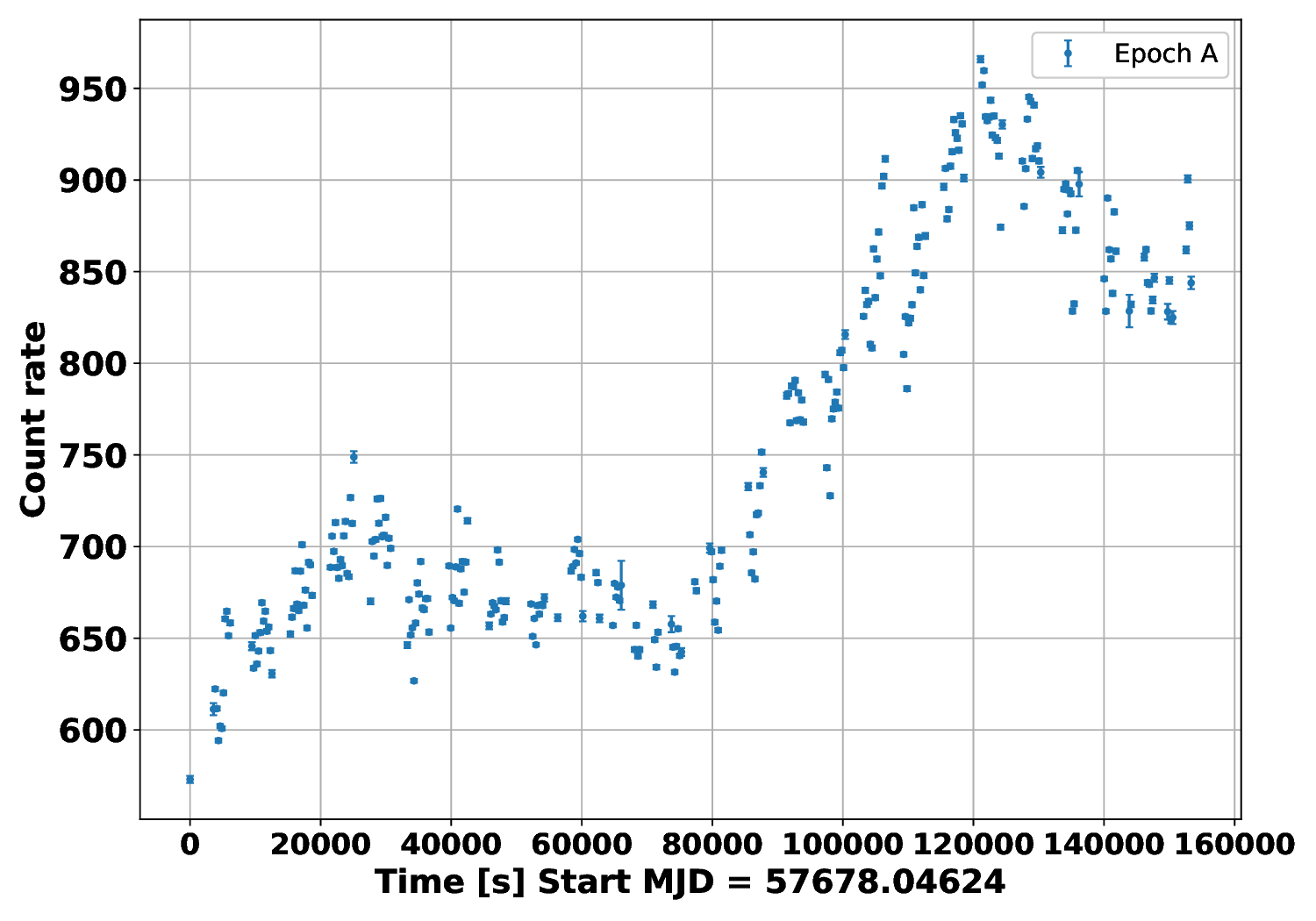}{0.45\textwidth}{}
          \fig{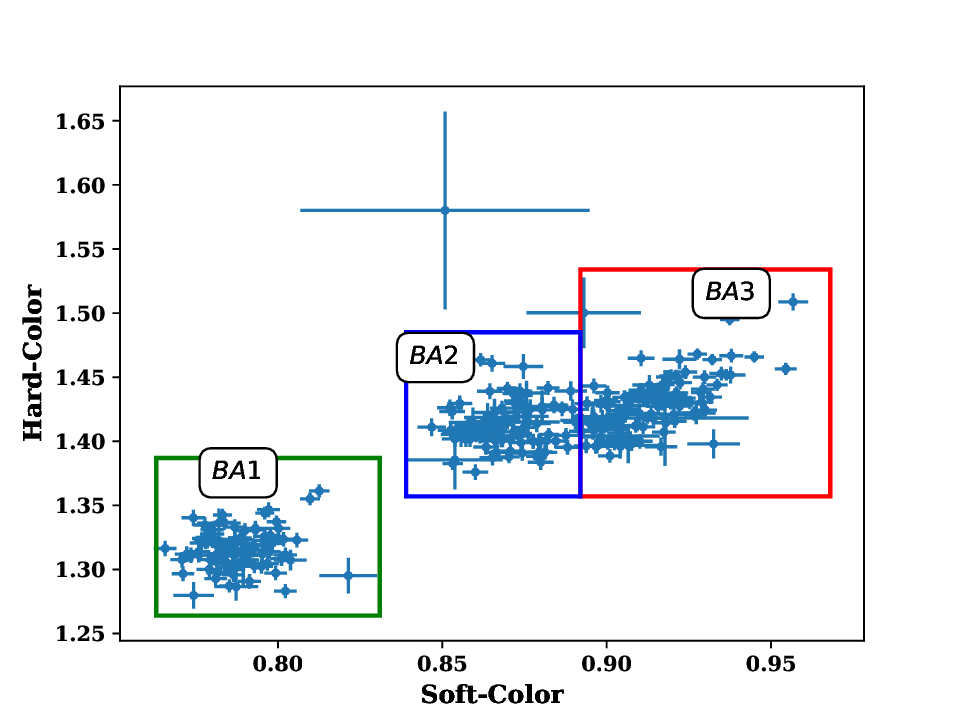}{0.45\textwidth}{}}
\gridline{\fig{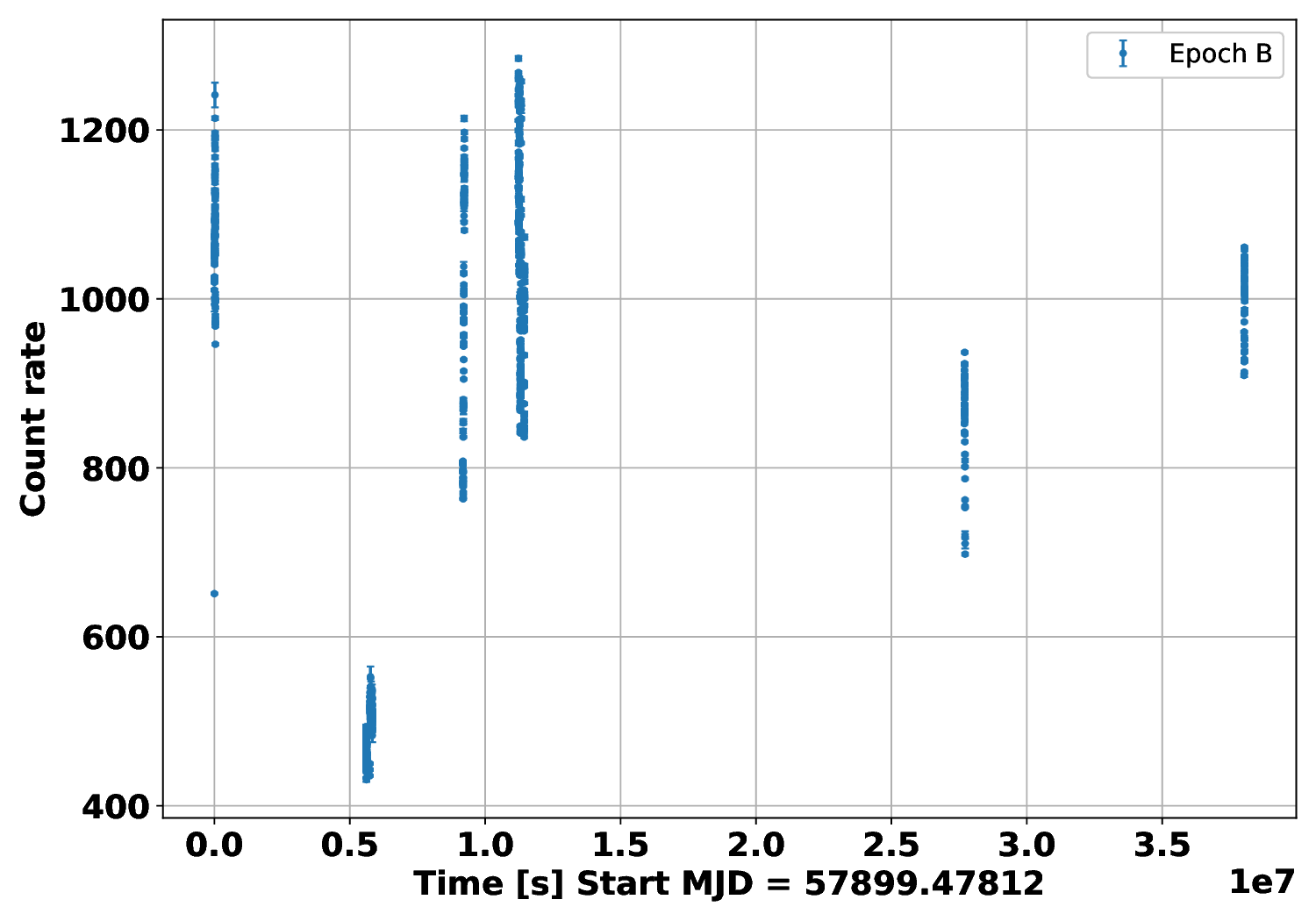}{0.45\textwidth}{}
          \fig{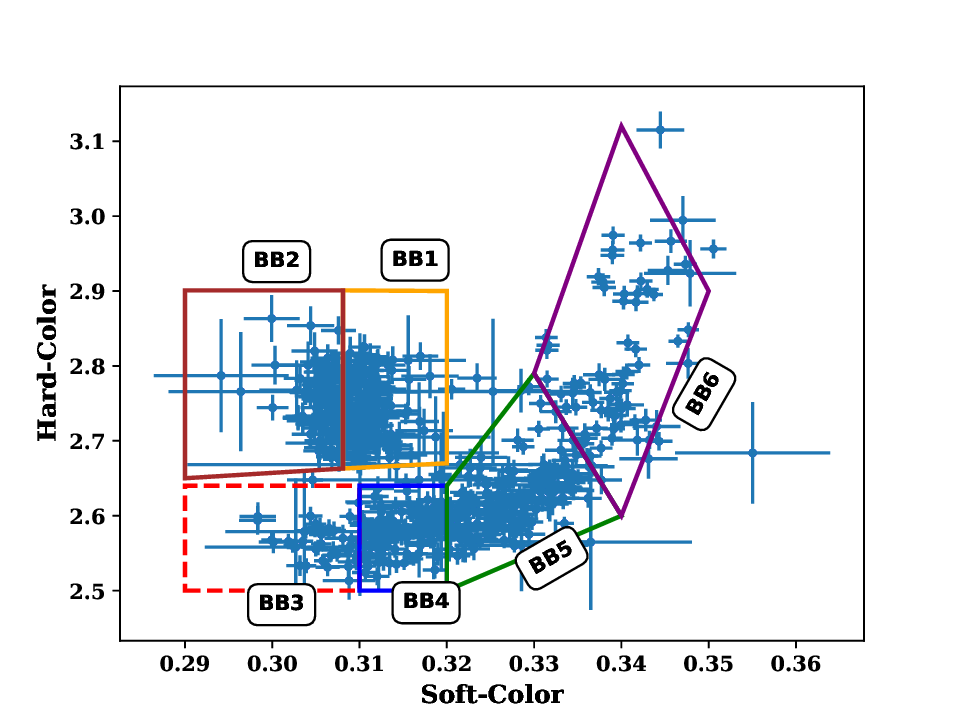}{0.45\textwidth}{}}
\gridline{\fig{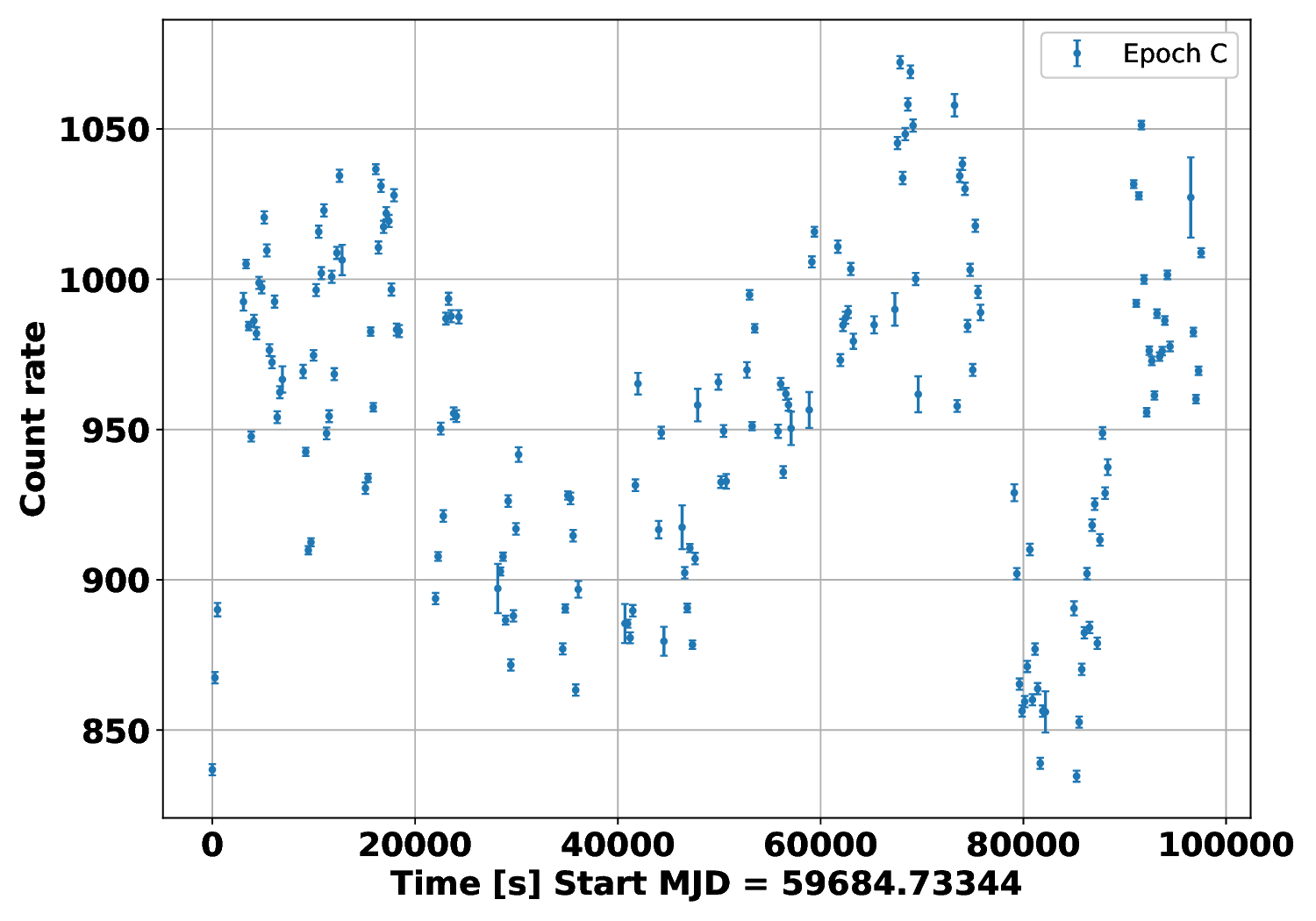}{0.45\textwidth}{}
          \fig{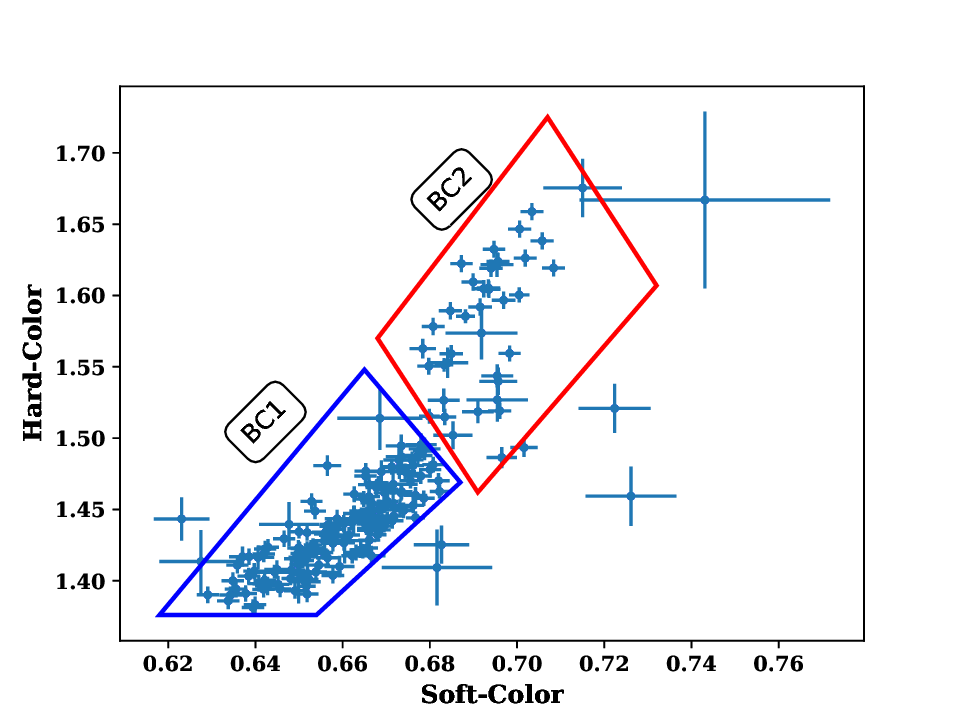}{0.45\textwidth}{}}
    \caption{The left column presents the 256-sec binned background-subtracted light curve within the energy range of 3.0-50.0 keV using LAXPC 20 data. In the right column, the CCD of 4U 1820-30 is constructed using three distinct energy bands: 3.0-5.0, 5.0-7.0, and 7.0-20.0 keV. Here the ratio of the count rates in the energy band 7.0-20.0 keV and 3.0-5.0 keV is defined as hard color, and the ratio of the count rates in the energy band 5.0-7.0 keV and 3.0-5.0 is defined as soft color. The light curve (left) and CCD (right) in the top panel corresponds to Epoch A, segmented into three parts: BA1, BA2, and BA3. The middle panel shows the light curve and CCD for Epoch B, divided into six segments named BB1, BB2, BB3, BB4, BB5, and BB6. The bottom panel is related to Epoch C, featuring two segments in the CCD, labelled BC1 and BC2. The highlighted areas indicate the portions used to generate the energy spectra and Power Density Spectra (PDS). Refer to the main text for more details.}
    \label{fig:lightcurve and CCD}
\end{figure}

We have plotted \textit{MAXI} count rate (2-20 keV) versus the hardness ratio to build the HID, as shown in Figure \ref{maxi_bat_hardness}. The hardness ratio is defined as the ratio between \textit{Swift/BAT} count rate (15-50 keV) and \textit{MAXI}\footnote{MAXI:\url{ http://maxi.riken.jp/top/index.html}, BAT: \url{https://swift.gsfc.nasa.gov/results/transients/}} count rate (2-4 keV).
The \textit{AstroSat} observations are highlighted as colored points. From Figure \ref{maxi_bat_hardness}, it is clear that all \textit{AstroSat} observations used in this work belong to the BS. We extracted the background-subtracted light curve from LAXPC20 event mode data in the 3.0-50.0 keV energy band for each epoch with a bin size of 256 sec. Epoch A has an average count rate of $\sim$ 754 cts/s, Epoch B  $\sim$ 842 cts/s, and Epoch C $\sim$ 956 cts/s. We have created the CCD using background-subtracted lightcurves in the 3.0-5.0 keV, 5.0-7.0 keV, and 7.0-20.0 keV energy bands. To show the CCD, we have plotted the hard color against the soft color. Here, the ratio of the count rates in the energy band 7.0-20.0 keV and 3.0-5.0 keV is defined as hard color, and the ratio of the count rates in the energy band 5.0-7.0 keV and 3.0-5.0 is defined as soft color. The light curves and CCD for the three Epochs are shown in Figure \ref{fig:lightcurve and CCD}. To study the spectral and temporal evolution along the ``\textbf{C}"-track, we have divided the track into 11 segments. BA1, BA2, and BA3 correspond to the Epoch A data, while Epoch B is divided into six segments: BB1, BB2, BB3, BB4, BB5, and BB6. Moreover, Epoch C is divided into two segments, BC1 and BC2. Each segment is marked in the CCD. For each segment, we have created the X-ray spectra and PDS, which are used for further spectral and temporal analysis.\\

\subsection{Timing Analysis}
 First, we have generated the light curves with a bin size of 0.0003 s in the 3.0-50.0 keV energy band. Light curves are divided into intervals of 8192 bins. For each interval, we created the PDS. The PDS corresponding to each segment of the CCD is averaged and then rebinned with a factor of 1.05. The binned PDS are normalized to obtain the fractional rms spectra in the units of $(rms/mean)^2 \,\mathrm{Hz}^{-1}$ \citep{Miyamoto}. Dead time-corrected Poisson noise level is subtracted from all the PDS \citep{zhang1995dead, agrawal20184U170544,Agarwa2020}. We have fitted the PDS in the frequency range 0.8-1500 Hz. The PDS is modeled with a combination of power-law and/or Lorentzian functions. Thus, we have found the presence of two types of noise features in the PDS: a VLFN modeled using a power-law function ($A\nu^{-\beta}$) with index $\beta$  and normalization $A$, and a BLN component modeled using the Lorentzian function defined as \citep{Belloni2002},

\begin{equation}
    P(\nu) = \frac{(LN) \nu_w}{2\pi} \frac{1}{\left [\left ( \nu - \nu_l \right )^2 + \left ( \nu_w/2 \right )^2  \right ]}
\end{equation}

\begin{figure}[h!]
    \centering
    \gridline{\rotatefig{360}{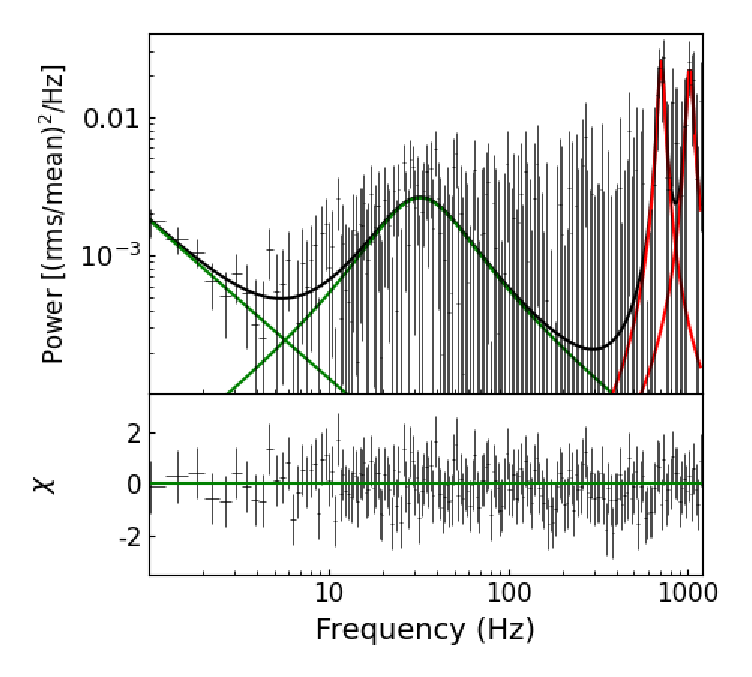}{0.50\textwidth}{}
              \rotatefig{360}{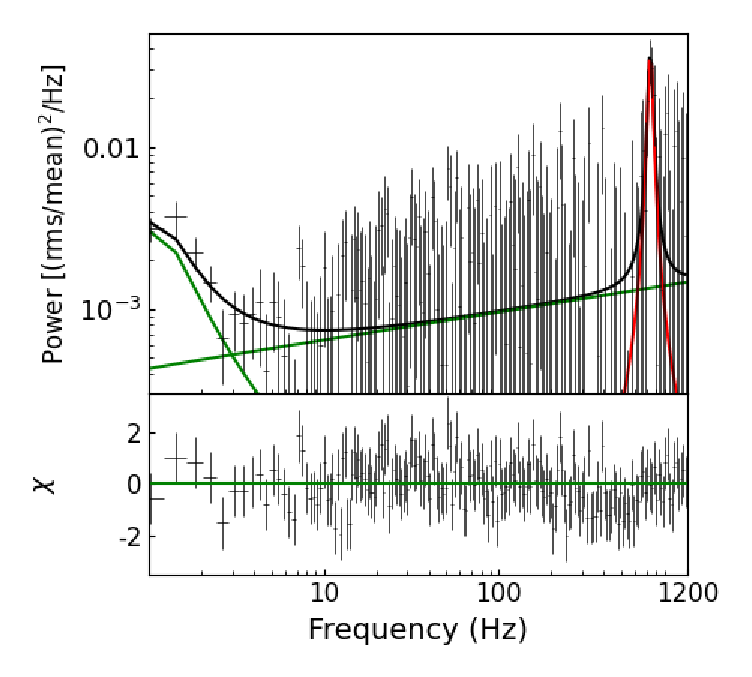}{0.50\textwidth}{}}
\caption{ The PDS for CDD segment BB1 (left panel) and BB2 (right panel) in the energy band 3.0-50.0 keV. The PDS shows significant kHz QPOs at $\sim$ 710 Hz and $\sim$ 740 Hz in the BB1 and BB2 segments. The residuals are shown in units of standard deviations from the model.}

    \label{PDS}
\end{figure}

\begin{table} [h]
    \centering
    \caption{The best fit parameters obtained by fitting the PDS in the 0.8-1500 Hz using 3.0-50.0 keV energy range for each segment.} 
    \begin{tabular}{cccccccccccc}
    \hline
        Segments & &VLFN & &LFN & HFN & QPO1 & QPO2& $\chi^2/dof$ & \\
        \hline
        \hline
 & $\beta$  & ${0.6}^{+0.1}_{-0.1}$ & ${\nu_L}$  & ${7.1}^{+2.0}_{-2.1}$ & &${}^{}_{}$ &  &\\
     BA1 & A ($\times 10^{-5}$)& ${11.8}^{+7.1}_{-4.6}$ & ${\nu_w}$ & ${30.8}^{+2.4}_{-2.4}$ & &${}^{}_{}$ &   & $136/189$\\
  &   &  &  $LN(\times 10^{-3})$ & ${13.1}^{+0.9}_{-0.9}$ & &${}^{}_{}$ &  &\\
  & & & rms $\%$ & ${11.4}^{+0.4}_{-0.4}$ & & &&\\
  \hline
 & $\beta$  & ${0.6}^{+0.1}_{-0.1}$ & ${\nu_L}$  & ${7.0}^{+2.1}_{-2.5}$ & &${}^{}_{}$ &  &\\
     BA2 & A ($\times 10^{-5}$) & ${7.1}^{+5.3}_{-3.1}$ & ${\nu_w}$ & ${30.5}^{+3.1}_{-3.1}$ & ${}^{}_{}$ &  & & $134/189$\\
  &   &  &  $LN (\times 10^{-3})$ & ${11.4}^{+0.8}_{-0.9}$ & ${}^{}_{}$ & &  &\\
  & & & rms $\%$ & ${10.7}^{+0.4}_{-0.4}$& & &&\\
  \hline

   & $\beta$  & ${0.2}^{+0.1}_{-0.1}$ & ${\nu_L}$  & ${7.9}^{+0.4}_{-0.5}$ & &${}^{}_{}$ & & \\
     BA3 & A ($\times 10^{-6}$) & ${2.0}^{+2.0}_{-1.0}$ & ${\nu_w}$ & ${16.6}^{+1.0}_{-0.9}$ & &${}^{}_{}$ &   & $111/187$\\
  &   &  &  $LN (\times 10^{-3})$ & ${12.6}^{+0.4}_{-0.4}$ & &${}^{}_{}$ &  & \\
  & & & rms $\%$ & ${11.2}^{+0.2}_{-0.2}$& & &&\\
  \hline

& $\beta$  & ${2.1}^{+0.4}_{-0.3}$ & ${\nu_L}$  & &${25.9}^{+4.1}_{-5.5}$&  ${710}^{+13}_{-14}$ & ${1026}^{+23}_{-20}$ &\\
      & A ($\times 10^{-3}$) & ${1.8}^{+0.3}_{-0.3}$ & ${\nu_w}$ & &${39}^{+14}_{-10}$  &${59}^{+27}_{-19}$ & $83^{+74}_{-48}$ & \\
BB1   &  &  &  $LN (\times 10^{-3})$ & &${4.3}^{+0.8}_{-0.8}$  &${3.3}^{+1.1}_{-0.9}$ & ${3.1}^{+1.5}_{-1.5}$  & $98/172$\\
   &    &   &  Q &  & & ${12}^{+6}_{-4}$ & ${12}^{+18}_{-6}$  &\\
   &    & &  rms (\%) & &${6.6}^{+0.6}_{-0.6}$  & ${5.7}^{+0.9}_{-0.8}$ & ${5.5}^{+1.2}_{-1.5}$   &\\
   &   &  &  $\sigma$ &  & & 3.7 & 2.4 &\\

\hline
& $\beta$  & ${0.9}^{+0.2}_{-0.1}$ & ${\nu_L}$  & ${0.90}^{+0.13}_{-0.12}$& & ${740}^{+8}_{-8}$ &  &\\
     BB2 & A ($\times 10^{-4}$) & ${4.5}^{+2.7}_{-1.7}$ & ${\nu_w}$ & ${1.0}^{+0.2}_{-0.2}$ & &${44}^{+23}_{-18}$ &  & $136/176$\\
  &   &  &  $LN (\times 10^{-3})$ & ${4.2}^{+0.8}_{-1.0}$ & &${3.8}^{+1.2}_{-1.1}$ &  &\\
   &   &  &  Q &  & & ${17}^{+12}_{-6}$ &   &\\
   &   &  &  rms (\%) & ${6.5}^{+0.6}_{-0.8}$ & & ${6.2}^{+0.9}_{-0.9}$ &   &\\
   &   &  &  $\sigma$ &  & & 3.7 &   &\\

  \hline
  & $\beta$  & ${0.3}^{+0.1}_{-0.1}$ & ${\nu_L}$  & ${6.3}^{+1.1}_{-1.2}$ & &${}^{}_{}$ &  &\\
     BB3 & A ($\times 10^{-6}$) & ${7.2}^{+6.5}_{-3.5}$ & ${\nu_w}$ & ${23}^{+2}_{-2}$ & &${}^{}_{}$ & & $146/189$\\
  &   &  &  $LN (\times 10^{-3})$ & ${11.3}^{+0.5}_{-0.5}$ & &${}^{}_{}$ & &\\
  & & & rms $\%$ & ${10.6}^{+0.3}_{-0.3}$ & & &&\\
  \hline

  & $\beta$  & ${0.2}^{+0.1}_{-0.1}$ & ${\nu_L}$  & ${6.8}^{+0.8}_{-0.8}$ & $25.3^{+1.5}_{-2.1}$ &${}^{}_{}$ &  &\\
     BB4 & A ($\times 10^{-6}$) & ${1.4}^{+1.2}_{-0.7}$ & ${\nu_w}$ & ${16.7}^{+2.2}_{-2.1}$ &$13.5^{+3.4}_{-3.1}$ &${}^{}_{}$ &   & $184/185$\\
  &   &  &  $LN (\times 10^{-3})$ & ${8.0}^{+0.8}_{-1.0}$ & $2.1^{+0.9}_{-0.7}$ &${}^{}_{}$ &  &\\
  & & & rms $\%$ &${8.9}^{+0.4}_{-0.5}$ & ${4.6}^{+0.9}_{-0.8}$ & &&\\
  \hline
  
&   & & ${\nu_L}$  & ${7.8}^{+0.4}_{-0.4}$ & &${}^{}_{}$ &  &\\
     BB5 &  & & ${\nu_w}$ & ${13.6}^{+0.7}_{-0.6}$ & &${}^{}_{}$ &   & $117/78$\\
  &   &  &  $LN (\times 10^{-3})$ & ${3.0}^{+0.2}_{-0.2}$ & &${}^{}_{}$ &  &\\
  & & & rms $\%$ & ${5.4}^{+0.2}_{-0.1}$& & &&\\
  \hline

&  &  & ${\nu_L}$  & & &${4.6}^{+0.2}_{-0.1}$ &   &\\
     BB6 &  &  & ${\nu_w}$ & & &${0.6}^{+0.8}_{-0.4}$ &    & $89/71$\\
  &   &  &  $LN (\times 10^{-4})$ & & & ${1.9}^{+0.9}_{-0.7}$ &   &\\
   & & & Q  & & &${7}^{+17}_{-4}$ & &\\
  & & & rms $\%$ &  & & ${1.4}^{+0.3}_{-0.3}$ & &\\
  & & &  $\sigma$ & & & 2.6 & &\\
  \hline

   & $\beta$  & ${0.1}^{+0.1}_{-0.1}$ & ${\nu_L}$  & ${7.7}^{+0.7}_{-0.4}$ & &${}^{}_{}$ &  &\\
     BC1 & A ($\times 10^{-6}$) & ${1.2}^{+2.0}_{-1.0}$ & ${\nu_w}$ & ${15.5}^{+1.6}_{-1.5}$ & &${}^{}_{}$ &   & $114/188$\\
  &   &  &  $LN (\times 10^{-3})$ & ${9.0}^{+0.5}_{-0.5}$ & &${}^{}_{}$ &  &\\
  & & & rms $\%$ &${9.5}^{+0.3}_{-0.3}$ & & &&\\
  \hline

 & $\beta$  & ${0.3}^{+0.6}_{-0.7}$ & ${\nu_L}$  & ${}^{}_{}$& & ${}^{}_{}$ &  &\\
     BC2 & A  ($\times 10^{-5}$) & ${0.3}^{+22.7}_{-0.3}$ & ${\nu_w}$ & ${}^{}_{}$& & ${}^{}_{}$ &   & $119/192$\\
  &   &  &  $LN (\times 10^{-3})$ & &${}^{}_{}$ & ${}^{}_{}$  & &\\
  & & & rms $\%$ & & & &&\\
  \hline
  \hline

    \end{tabular}
    \label{tab:Timing}
\end{table}

where LN is the normalization, $\nu_l$ is the centroid frequency, and $\nu_w$  is the full width at half maximum (FWHM). In this representation, the square root of the normalization (LN) gives the fractional rms amplitude.
The ratio between $\nu_l$ and  $\nu_w$ is called the quality factor (Q) of the Lorentzian feature. A narrow feature with Q $>3$ is called a QPO \citep{Belloni2002,Agarwa2020}. We consider only those  QPOs detected at a significance level greater than 3$\sigma$. In Figure \ref{PDS}, we have shown the PDS for BB1 and BB2 segments. However, the best-fit values of centroid frequency, FWHM, Q-value, sigma value, and percentage fractional rms amplitude for all segments are reported in Table \ref{tab:Timing}. 1$\sigma$ errors in the best-fit parameters are calculated using $\Delta \chi^2=1$.

\subsection{Spectral Analysis}
\label{spectral_analysis}
For every segment, we have produced the background and source spectra. All spectra are further optimally grouped using \texttt{ftgrouppha}. The grouped spectra were fitted simultaneously using the spectral fitting package \texttt{XSPEC version 12.12.1} \citep{arnaud1996astronomical} in the energy range of 0.7-20 keV. For SXT, we have utilized an energy range of 0.7-7 keV and  5–20 keV for LAXPC20, ignoring data beyond 20 keV because of background domination. To account for the response matrix's uncertainty, we added 3\% systematic during the fitting process \citep{Systematic}. We have computed the error at 1$\sigma$ (68\% confidence level). 
 \begin{figure} [h!]
    \centering
    \gridline{\fig{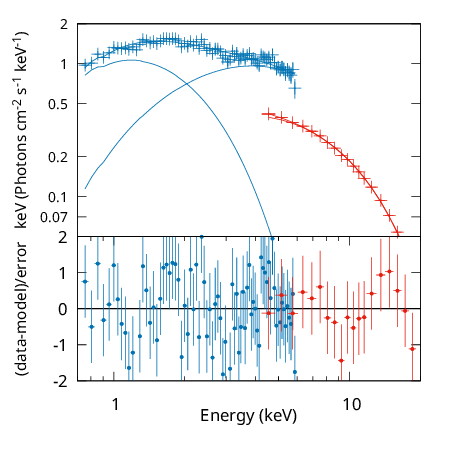}{0.50\textwidth}{}
              \fig{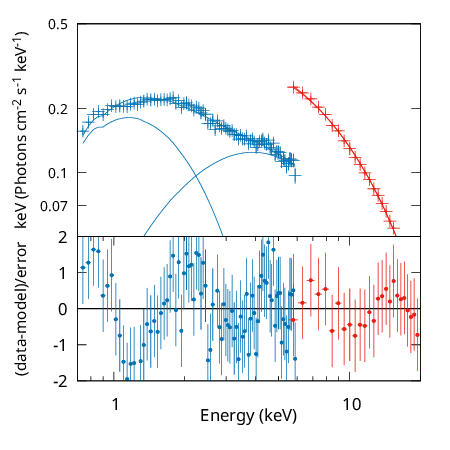}{0.50\textwidth}{}}
\caption{SXT (blue, 0.7-7.0 keV) and LAXPC20 (red, 5.0-20.0 keV) unfolded spectra of Segment BA1 (left panel) and BB1 (right panel), fit with the Model 3 - \texttt{constant*tbabs*(Comptb+diskbb)}. The residuals are shown in units of standard deviations from the model}
    \label{fig:spectra}
\end{figure}

\begin{table}[h]
 \caption{Best fit parameters obtained by fitting the spectrum of each segment on CCD for each Epoch using Model 1: \texttt{constant* TBabs*(diskbb+bbodyrad)}. The best fit parameters are the disk temperature $kT_{\mathrm{in}}$ (unit in keV), disk normalization $N_{\mathrm{dbb}}$, black body temperature $kT_{\mathrm{bb}}$ (unit in keV), and black body normalization $N_{\mathrm{BB}}$. The reduced $\chi^2(\chi^2/dof)$ is also reported in the last column of this table.}
\label{Table bbodyrad_diskbb}
\begin{tabular}{ccccccc}
\hline
&Segments & $kT_{\mathrm{in}}$& $N_{\mathrm{dbb}}$ & $kT_{\mathrm{bb}}$ & $N_{\mathrm{BB}}$ & $\chi^2/dof$ \\ \hline\hline
&BA &  $1.13_{-0.02}^{+0.02}$ & $260_{-20}^{+20}$ & $2.26_{-0.01}^{+0.01}$ & $34.9_{-1.5}^{+1.5}$ & 132/92 \\
Epoch A &BA2  &  $1.15_{-0.01}^{+0.01}$ & $270_{-10}^{+10}$ & $2.36_{-0.01}^{+0.01}$ & $23.8_{-1.3}^{+1.3}$ & 142/101 \\
&BA3 &  $1.38_{-0.03}^{+0.03}$ & $70_{-4}^{+5}$ & $2.42_{-0.02}^{+0.02}$ & $14.1_{-0.9}^{+0.9}$ & 93/107 \\
\hline
&BB1  &  $1.57_{-0.03}^{+0.03}$ & $50_{-3}^{+4}$ & $2.71_{-0.04}^{+0.04}$ & $6.0_{-0.6}^{+0.7}$ & 108/92 \\
&BB2  &  $1.64_{-0.03}^{+0.03}$ & $50_{-3}^{+3}$ & $2.74_{-0.04}^{+0.04}$ & $6.4_{-0.6}^{+0.6}$ & 145/102 \\
Epoch B&BB3  &  $1.74_{-0.03}^{+0.03}$ & $20_{-1}^{+2}$ & $2.81_{-0.03}^{+0.03}$ & $3.2_{-0.3}^{+0.3}$ & 172/100 \\
&BB4  &  $1.47_{-0.03}^{+0.03}$ & $70_{-5}^{+5}$ & $2.59_{-0.03}^{+0.03}$ & $13.6_{-1.0}^{+1.0}$ & 167/101 \\
&BB5  &  $1.32_{-0.01}^{+0.01}$ & $60_{-2}^{+2}$ & $2.65_{-0.02}^{+0.02}$ & $4.8_{-0.2}^{+0.2}$ & 186/107 \\
&BB6  &  $1.31_{-0.01}^{+0.01}$ & $60_{-2}^{+3}$ & $2.64_{-0.02}^{+0.02}$ & $4.9_{-0.2}^{+0.2}$ & 157/107 \\
\hline
Epoch C&BC1&  $1.44_{-0.02}^{+0.02}$ & $84_{-4}^{+5}$ & $2.34_{-0.03}^{+0.03}$ & $12.5_{-1.2}^{+1.2}$ & 175/100 \\
&BC2  &  $1.44_{-0.03}^{+0.03}$ & $74_{-5}^{+5}$ & $2.37_{-0.02}^{+0.02}$ & $14.6_{-1.3}^{+1.3}$ & 107/96 \\
\hline
\end{tabular}

\end{table}
We have attempted multi-component phenomenological spectral models to fit the spectra. Initially, we modeled the spectrum of each segment using Model 1: \texttt{constant*TBabs*(diskbb+bbodyrad)}. Here, \texttt{constant} is used to handle the relative normalization between the two instruments: SXT and LAXPC20. \texttt{TBabs} is an absorption model with the abundances given by \cite{Wilims} and the cross-section as in \cite{Verner}. For our entire analysis, we fixed the hydrogen column density ($N_H$) value at $\sim 0.14 \times 10^{22} \mathrm{cm^{-2}}$ \citep{IXPE2023}.
We applied the \texttt{gain fit} command to adjust the gain shift in the SXT.   
Our Model 1 is the combination of a blackbody emission component from the BL/NS surface (\texttt{bbodyrad} in \texttt{XSPEC}) and a MCD model (\texttt{diskbb} in \texttt{XSPEC}) \citep{Mistuda, Makishima}. The best-fit parameters are reported in the Table \ref{Table bbodyrad_diskbb}.
Then, we replaced the \texttt{bbodyrad} with \texttt{nthComp}. The \texttt{nthComp} model takes care of the emission from the thermally Comptonized corona  \citep{Zdziarski1996}. The model becomes Model 2: \texttt{constant*TBabs*(nthcomp+diskbb)}. The best-fit parameters for the segments corresponding to Epoch A and Epoch C  are tabulated in Table  \ref{Table Epoch AC}. For Epoch B, all parameters are reported in Table \ref{table Epoch B}. The spectra could be well described by the MCD model along with the thermally Comptonized emission model from the corona, which is confirmed by the improved $\chi^2$ obtained by the Model 2 as compare to Model 1. To gain deeper insight into the physical properties of the BL/corona, we further replaced \texttt{nthcomp} with the more physically motivated model \texttt{comptb}, and model becomes Model 3:  \texttt{constant*TBabs*(comptb+diskbb)}. Here, \texttt{comptb} model comprises two components: the direct soft seed photons that are not up-scattered, and the Comptonized emission \citep{comptb}. During this fitting, the blackbody seed photon index ($\gamma$) and bulk efficiency ($\delta$) are kept fixed at 3 and 0, respectively. Since the logarithm of the illumination parameter (log A) saturated at 8, we kept it fixed at this value. Although both Model 2 and Model 3 provides comparable reduced $\chi^2$ values, we consider Model 3 for all future considerations and estimation of the physical parameters because it offers insight into the BL geometry and inflow dynamics near the NS surface. Figure \ref{fig:spectra} shows the unfolded spectra along with the residuals for segments BA1 and BB1. The best-fit parameters obtained using Model 3 are listed in Table \ref{table comptb AC} and Table \ref{table comptb B}. We did not observe Fe emission line and reflection features, or Compton hump from any segment.

\begin{table}[h]
\caption{Best fit parameters obtained by fitting the spectrum of each segment on CCD for Epoch A and Epoch C using Model 2: \texttt{constant*TBabs*(nthcomp+diskbb)}. Here, $\Gamma$: photon index, $kT_{\mathrm{e}}$: electron temperature, $kT_{\mathrm{bb}}$: seed photon temperature, $N_{comp}$: normalization of the Comptonized component, $\tau$: optical thickness; $kT_{\mathrm{in}}$: disk temperature, $N_{\mathrm{dbb}}$: disk normalization, $R_{\mathrm{in}}$: inner disk radius in km, $\dot{m}$: mass accretion rate in unit of $\mathrm{g\; cm^{-2}\; s^{-1}}$. $F_{\mathrm{bol}}$, $F_{\mathrm{tot}}$, and $F_{\mathrm{disk}}$ are the unabsorbed bolometric flux in 0.01-200 keV energy range, total unabsorbed flux and unabsorbed disk flux in 0.7-20.0 keV energy range respectively, in the unit of $\mathrm{erg \;cm^{-2} \;s^{-1}}$. $\mu$: magnetic dipole moment, B: magnetic field strength, ($R_{\mathrm{max}}-R_{*}$): thickness of BL and $R_{\mathrm{max}}$: radial extent of the BL.}
\label{Table Epoch AC}
\begin{tabular}{ccccc|cc}
\hline
 & & Epoch A & &  & Epoch C & \\ \hline
component & parameter & BA1 & BA2 & BA3 & BC1 & BC2  \\ \hline
    nthcomp  & $\Gamma$ &  $1.78_{-0.04}^{+0.04}$ & $1.71_{-0.03}^{+0.03}$ & $1.71_{-0.05}^{+0.05}$ & $2.09_{-0.03}^{+0.03}$ & $2.01_{-0.04}^{+0.04}$ \\
 & $kT_e$ (keV) &  $2.68_{-0.05}^{+0.05}$ & $2.78_{-0.05}^{+0.05}$ & $2.75_{-0.06}^{+0.05}$ & $2.84_{-0.06}^{+0.05}$ & $2.89_{-0.07}^{+0.06}$ \\
 & $kT_{bb}$ (keV) & $0.98_{-0.03}^{+0.03}$ & $0.95_{-0.03}^{+0.03}$ & $0.96_{-0.04}^{+0.04}$ & $0.91_{-0.03}^{+0.03}$ & $1.08_{-0.05}^{+0.06}$\\
 & ${N_{comp}}$ &  $0.30_{-0.01}^{+0.01}$ & $0.32_{-0.01}^{+0.01}$ & $0.16_{-0.01}^{+0.01}$ & $0.27_{-0.01}^{+0.01}$ &  $0.17_{-0.01}^{+0.01}$ \\
& $\tau$ &  $12.5_{-0.5}^{+0.5}$ & $13.0_{-0.5}^{+0.5}$ & $13.1_{-0.1}^{+0.1}$ & $9.6_{-0.3}^{+0.3}$ & $10.0_{-0.4}^{+0.4}$\\
& $y-par$ & $3.3_{-0.2}^{+0.3}$ & $3.7_{-0.2}^{+0.4}$ & $3.7_{-0.1}^{+0.1}$  & $2.0_{-0.1}^{+0.2}$ &$2.3_{-0.1}^{+0.2}$ \\
  diskbb & $kT_{\mathrm{in}}$ (keV) &  $0.64_{-0.02}^{+0.02}$ & $0.71_{-0.02}^{+0.02}$ & $0.74_{-0.03}^{+0.03}$ & $0.52_{-0.01}^{+0.01}$  & $0.62_{-0.01}^{+0.01}$ \\
& $N_{\mathrm{dbb}}$ & $1520_{-180}^{+190}$& $1090_{-100}^{+110}$ & $370_{-50}^{+60}$ & $1710_{-150}^{+170}$& $830_{-80}^{+90}$ \\
& $R_{\mathrm{in}}$ (km) &  $36_{-2}^{+2}$  & $30_{-2}^{+2}$ &  $18_{-1}^{+1}$ &  $38_{-2}^{+2}$ & $26_{-1}^{+1}$ \\
& $\dot{m} (\times 10^4 $ $  \mathrm{g\; cm^{-2}\; s^{-1}})$&  $8.92_{-0.04}^{+0.04}$ & $9.96_{-0.03}^{+0.03}$ & $4.81_{-0.01}^{+0.01}$ & $5.24_{-0.01}^{+0.01}$ & $5.22_{-0.02}^{+0.02}$\\
\hline
& $F_{\mathrm{bol}}(\times 10^{-9} \mathrm{erg \;cm^{-2} \;s^{-1}})$ & $19.61_{-0.09}^{+0.09}$ & $21.89_{-0.08}^{+0.08}$ & $10.57_{-0.04}^{+0.04}$ & $11.51_{-0.04}^{+0.04}$ & $11.48_{-0.04}^{+0.04}$ \\
 & $F_{\mathrm{tot}}(\times 10^{-9} \mathrm{erg \;cm^{-2} \;s^{-1}})$ &  $17.63_{-0.08}^{+0.08}$ & $19.81_{-0.07}^{+0.07}$ & $9.72_{-0.03}^{+0.03}$ & $10.31_{-0.03}^{+0.03}$ & $10.42_{-0.04}^{+0.04}$ \\
  & $F_{\mathrm{nthcomp}}(\times 10^{-9} \mathrm{erg \;cm^{-2} \;s^{-1}})$ &  $13.68_{-0.08}^{+0.08}$ & $15.37_{-0.07}^{+0.07}$ &  $7.84_{-0.04}^{+0.04}$ & $8.60_{-0.03}^{+0.03}$ & $8.45_{-0.04}^{+0.04}$ \\
  & $F_{\mathrm{disk}}(\times 10^{-9} \mathrm{erg \;cm^{-2} \;s^{-1}})$ &  $3.92_{-0.04}^{+0.04}$ & $4.49_{-0.03}^{+0.03}$ & $1.87_{-0.01}^{+0.01}$& $1.70_{-0.01}^{+0.01}$ & $1.96_{-0.01}^{+0.02}$ \\
& $F_{nthcomp}/F_{tot}$ & 0.78 & 0.77 & 0.80 & 0.83 & 0.81 \\
 & $F_{disk}/F_{tot}$ & 0.22 & 0.23 & 0.20 & 0.17 & 0.19 \\
& $\mu (\times 10^{26}\mathrm{G \; cm^3})$ & 19 & 15 & 4 & 15 & 8 \\  
 & B ($\times 10^{8}$G)& 50 & 38 & 11 & 41 & 22 \\ 
  & ($R_{max}-R_{*}$) (km) & 25 & 29 & 11 & 13 & 13 \\
 & $R_{max}$ (km) & 40 & 38 & 20 & 22 & 22 \\ 
\hline
 & $\chi^2/dof$ &  79.22/91 & 81.57/100 & 84.28/106  & 83.33/100 & 70.94/98 \\
        \hline
\end{tabular}

\end{table}

\begin{table}
\caption{Best fit parameters obtained by fitting the spectrum of each segment on CCD for Epoch B using Model 2: \texttt{constant*TBabs*(nthcomp+diskbb)}. Here, $\Gamma$: photon index, $kT_{\mathrm{e}}$: electron temperature, $kT_{\mathrm{bb}}$: seed photon temperature, $N_{comp}$: normalisation of the Comptonized component, $\tau$: optical thickness, $kT_{\mathrm{in}}$: disk temperature, $N_{\mathrm{dbb}}$: disk normalization, $R_{\mathrm{in}}$: inner disk radius in km, $\dot{m}$: mass accretion rate in unit of $\mathrm{g\; cm^{-2}\; s^{-1}}$. $F_{\mathrm{bol}}$, $F_{\mathrm{tot}}$, and $F_{\mathrm{disk}}$ are the unabsorbed bolometric flux in 0.01-200 keV energy range, total unabsorbed flux and unabsorbed disk flux in 0.7-20.0 keV energy range respectively, in the unit of $\mathrm{erg \;cm^{-2} \;s^{-1}}$. $\mu$: magnetic dipole moment, B: magnetic field strength, ($R_{max}-R_{*}$): thickness of BL and $R_{max}$: radial extent of the BL.}
\label{table Epoch B}
\begin{tabular}{cccccccc}
\hline
component & parameter & BB1 & BB2 & BB3 & BB4 & BB5 & BB6 \\ \hline
    nthcomp  & $\Gamma$ & $2.08_{-0.11}^{+0.08}$ & $2.00_{-0.08}^{+0.09}$ & $2.30_{-0.11}^{+0.29}$ & $2.40_{-0.13} ^{+0.18}$ & $2.30_{-0.15}^{+0.15}$ & $2.11_{-0.14}^{+0.11}$  \\
        & $kT_e$ (keV) & $3.64_{-0.21}^{+0.18}$ & $3.48_{-0.14}^{+0.18}$ & $3.93_{-0.30}^{+1.02}$ & $4.20_{-0.37}^{+0.68}$ & $3.95_{-0.35}^{+0.42}$ & $3.71_{-0.30}^{+0.35}$  \\
        & $kT_{\mathrm{bb}}$ (keV) & $1.08_{-0.11}^{+0.06}$ & $1.01_{-0.10}^{+0.05}$ &  $1.08_{-0.07}^{+0.14}$ & $1.16_{-0.06}^{+0.06}$ & $1.12_{-0.08}^{+0.08}$ & $1.03_{-0.11}^{+0.18}$  \\
        & $N_\mathrm{comp}$ & $0.08_{-0.01}^{+0.01}$ & $0.09_{-0.01}^{+0.01}$ &  $0.16_{-0.03}^{+0.03}$ & $0.15_{-0.01}^{+0.01}$ & $0.10_{-0.01}^{+0.01}$ & $0.19_{-0.01}^{+0.01}$  \\
         & $\tau$ & $8.4_{-0.6}^{+1.0}$ & $9.1_{-0.8}^{+0.8}$ &  $7.0_{-1.8}^{+0.9}$ & $6.3_{-1.0}^{+0.9}$ & $7.0_{-0.9}^{+1.1}$ & $8.1_{-0.9}^{+1.2}$  \\
          & $y-par$ & $1.99_{-0.17}^{+0.63}$ & $2.23_{-0.27}^{+0.56}$ &  $1.49_{-0.69}^{+0.90}$ & $1.32_{-0.29}^{+0.68}$ & $1.49_{-0.26}^{+0.72}$ & $1.91_{-0.27}^{+0.88}$ \\
    diskbb & $kT_{\mathrm{in}}$ (keV) & $0.62_{-0.03}^{+0.01}$ & $0.62_{-0.03}^{+0.01}$  &  $0.68_{-0.02}^{+0.07}$ & $0.61_{-0.02}^{+0.02}$ & $0.64_{-0.04}^{+0.05}$ & $0.63_{-0.05}^{+0.03}$ \\
        & $N_{\mathrm{dbb}}$ & $640_{-60}^{+130}$ & $630_{-50}^{+120}$ &  $570_{-140}^{+110}$ & $850_{-110}^{+90}$ & $370_{-70}^{+80}$ & $810_{-70}^{+240}$ \\
        & $R_{\mathrm{in}}$ (km) & $23_{-1}^{+2}$ & $23_{-1}^{+2}$ &  $22_{-3}^{+2}$ & $27_{-2}^{+1}$ & $17_{-2}^{+2}$ & $26_{-1}^{+4}$  \\
        & $\dot{m} (\times 10^4 $ $  \mathrm{g\; cm^{-2}\; s^{-1}})$ & $2.94_{-0.01}^{+0.01}$ & $2.95_{-0.01}^{+0.01}$ &  $4.63_{-0.02}^{+0.02}$ & $4.94_{-0.01}^{+0.01}$ & $2.99_{-0.01}^{+0.01}$ & $6.29_{-0.02}^{+0.02}$ \\
     \hline
        & $F_{\mathrm{bol}}(\times 10^{-9} \mathrm{erg \;cm^{-2} \;s^{-1}})$ & $6.46_{-0.02}^{+0.02}$ & $6.49_{-0.02}^{+0.02}$&  $10.18_{-0.04}^{+0.04}$ & $10.87_{-0.04}^{+0.04}$ & $6.57_{-0.02}^{+0.02}$ & $13.83_{-0.05}^{+0.05}$ \\
        & $F_{\mathrm{Tot}}(\times 10^{-9} \mathrm{erg \;cm^{-2} \;s^{-1}})$ & $5.61_{-0.01}^{+0.01}$ & $5.66_{-0.02}^{+0.02}$&  $9.09_{-0.04}^{+0.04}$ & $9.64_{-0.03}^{+0.03}$ & $5.82_{-0.02}^{+0.02}$ & $12.42_{-0.04}^{+0.04}$  \\
        & $F_{\mathrm{nthcomp}}(\times 10^{-9} \mathrm{erg \;cm^{-2} \;s^{-1}})$ & $4.16_{-0.01}^{+0.01}$ & $4.21_{-0.02}^{+0.02}$ &  $7.11_{-0.04}^{+0.04}$ & $7.77_{-0.03}^{+0.03}$ & $4.82_{-0.02}^{+0.02}$ & $10.41_{-0.05}^{+0.05}$  \\
        & $F_{\mathrm{disk}}(\times 10^{-9} \mathrm{erg \;cm^{-2} \;s^{-1}})$  &$1.45_{-0.01}^{+0.01}$ & $1.43_{-0.01}^{+0.01}$ &  $1.98_{-0.02}^{+0.02}$ & $1.87_{-0.01}^{+0.01}$ & $0.99_{-0.01}^{+0.01}$ & $2.01_{-0.01}^{+0.01}$   \\
       
        & $F_{nthcomp}/F_{tot}$ & 0.75 & 0.75 &  0.78 & 0.81 & 0.83 & 0.84  \\
         & $F_{disk}/F_{tot}$ & 0.25 & 0.25 &  0.22 & 0.19 & 0.17 & 0.16   \\

          & $\mu (\times 10^{26}\mathrm{G \; cm^3})$ & 5 & 5 &  6 & 8 & 3 & 10  \\
          & B ($\times 10^{8}$G)& 14 & 14 &  16 & 22 & 9 & 27  \\
          & ($R_{max}-R_{*}$) (km) & 7 & 7 & 11 & 12 & 7 & 16  \\ 
        & $R_{max}$ (km) & 16 & 16 & 20 & 21 & 16 & 25  \\ 
         \hline
        & $\chi^2/dof$ & 87.39/105 & 101.44/105 &  94.67/90 & 78.78/100 & 63.82/99 & 65.39/99  \\
        \hline
\end{tabular}
\end{table}

\begin{table}[h]
\caption{Best fit parameters obtained by fitting the spectrum of each segment on CCD for Epoch A and Epoch C using Model 3: \texttt{constant*TBabs*(comptb+diskbb)}. Here, $\Gamma$: photon index, $kT_{\mathrm{e}}$: electron temperature, $kT_{\mathrm{s}}$: seed photon temperature, $N_{comp}$: normalisation of the Comptonized component, $\tau$: optical thickness, $kT_{\mathrm{in}}$: disk temperature, $N_{\mathrm{dbb}}$: disk normalization, $R_{\mathrm{in}}$: inner disk radius in km, $\dot{m}$: mass accretion rate in unit of $\mathrm{g\; cm^{-2}\; s^{-1}}$. $F_{\mathrm{bol}}$, $F_{\mathrm{tot}}$, and $F_{\mathrm{disk}}$ are the unabsorbed bolometric flux in 0.01-200 keV energy range, total unabsorbed flux and unabsorbed disk flux in 0.7-20.0 keV energy range respectively, in the unit of $\mathrm{erg \;cm^{-2} \;s^{-1}}$. $\mu$: magnetic dipole moment, B: magnetic field strength, ($R_{\mathrm{max}}-R_{*}$): thickness of BL and $R_{\mathrm{max}}$: radial extent of the BL.}
\label{table comptb AC}
\begin{tabular}{ccccc|cc}
\hline
 & & Epoch A & &  & Epoch C & \\ \hline
component & parameter & BA1 & BA2 & BA3 & BC1 & BC2  \\ \hline
    comptb  & $\Gamma = (1-\alpha)$ &  $1.76_{-0.08}^{+0.14}$ & $1.74_{-0.06}^{+0.10}$ & $1.72_{-0.06}^{+0.08}$ & $2.15_{-0.09}^{+0.11}$ & $1.92_{-0.08}^{+0.11}$ \\
 & $kT_e$ (keV) &  $2.58_{-0.06}^{+0.09}$ & $2.73_{-0.06}^{+0.08}$ & $2.68_{-0.05}^{+0.06}$ & $2.77_{-0.09}^{+0.11}$ & $2.70_{-0.08}^{+0.08}$ \\
 & $kT_{s}$ (keV) & $0.91_{-0.26}^{+0.22}$ & $0.91_{-0.17}^{+0.15}$ & $0.87_{-0.14}^{+0.15}$ & $0.94_{-0.06}^{+0.06}$ & $0.94_{-0.10}^{+0.10}$\\
 & ${N_{comptb}} (\times 10^{-2})$ &  $9.7_{-1.0}^{+1.0}$ & $10.4_{-0.8}^{+0.8}$ & $5.4_{-0.4}^{+0.3}$ & $6.8_{-0.2}^{+0.2}$ &  $6.6_{-0.3}^{+0.3}$ \\
& $\tau$ &  $13.0_{-1.6}^{+1.2}$ & $12.8_{-1.2}^{+0.9}$ & $13.2_{-1.1}^{+0.9}$ & $9.4_{-0.8}^{+0.8}$ & $11.1_{-1.1}^{+0.9}$\\
& $y-par$ & $3.4_{-0.8}^{+0.8}$ & $3.5_{-0.6}^{+0.6}$ & $3.6_{-0.5}^{+0.6}$  & $1.9_{-0.3}^{+0.4}$ &$2.6_{-0.4}^{+0.5}$ \\
  diskbb & $kT_{\mathrm{in}}$ (keV) &  $0.64_{-0.12}^{+0.07}$ & $0.60_{-0.06}^{+0.05}$ & $0.58_{-0.06}^{+0.06}$ & $0.54_{-0.03}^{+0.03}$  & $0.56_{-0.04}^{+0.04}$ \\
& $N_{\mathrm{dbb}}$ & $1460_{-400}^{+1400}$& $1900_{-400}^{+700}$ & $950_{-240}^{+400}$ & $1450_{-240}^{+300}$& $1130_{-230}^{+320}$ \\
& $R_{\mathrm{in}}$ (km) &  $35_{-5}^{+14}$  & $40_{-4}^{+8}$ &  $28_{-4}^{+6}$ &  $35_{-3}^{+3}$ & $31_{-3}^{+4}$ \\
& $\dot{m} (\times 10^4 $ $  \mathrm{g\; cm^{-2}\; s^{-1}})$&  $8.90_{-0.04}^{+0.04}$ & $9.82_{-0.03}^{+0.03}$ & $4.89_{-0.01}^{+0.01}$ & $5.20_{-0.01}^{+0.01}$ & $5.25_{-0.02}^{+0.02}$\\
\hline
& $F_{\mathrm{bol}}(\times 10^{-9} \mathrm{erg \;cm^{-2} \;s^{-1}})$ & $19.56_{-0.09}^{+0.09}$ & $21.59_{-0.07}^{+0.07}$ & $10.72_{-0.04}^{+0.04}$ & $11.44_{-0.04}^{+0.04}$ & $11.56_{-0.04}^{+0.04}$ \\
 & $F_{\mathrm{tot}}(\times 10^{-9} \mathrm{erg \;cm^{-2} \;s^{-1}})$ &  $17.63_{-0.08}^{+0.08}$ & $19.33_{-0.07}^{+0.07}$ & $9.70_{-0.03}^{+0.03}$ & $10.28_{-0.03}^{+0.03}$ & $10.47_{-0.04}^{+0.04}$ \\
  & $F_{\mathrm{comptb}}(\times 10^{-9} \mathrm{erg \;cm^{-2} \;s^{-1}})$ &  $13.84_{-0.08}^{+0.08}$ & $15.42_{-0.07}^{+0.07}$ &  $8.10_{-0.04}^{+0.04}$ & $8.50_{-0.03}^{+0.03}$ & $8.76_{-0.04}^{+0.04}$ \\
  & $F_{\mathrm{disk}}(\times 10^{-9} \mathrm{erg \;cm^{-2} \;s^{-1}})$ &  $3.78_{-0.04}^{+0.04}$ & $3.91_{-0.03}^{+0.03}$ & $1.60_{-0.01}^{+0.01}$& $1.78_{-0.01}^{+0.01}$ & $1.70_{-0.01}^{+0.01}$ \\
& $F_{comptb}/F_{tot}$ & 0.78 & 0.80 & 0.83 & 0.83 & 0.84 \\
 & $F_{disk}/F_{tot}$ & 0.22 & 0.20 & 0.17 & 0.17 & 0.16 \\
& $\mu (\times 10^{26}\mathrm{G \; cm^3})$ & 29 & 29 & 11 & 15 & 12 \\  
 & B ($\times 10^{8}$G)& 77 & 75 & 30 & 39 & 33 \\ 
  & ($R_{max}-R_{*}$) (km) & 25 & 28 & 12 & 12 & 13 \\
 & $R_{max}$ (km) & 34 & 38 & 21 & 22 & 22 \\ 
  & $R_{m}$ (km) & 38 & 37 & 26 & 30 & 27 \\ 
\hline
 & $\chi^2/dof$ &  80.56/89 & 94.09/99 & 108.03/105  & 82.61/99 & 69.30/97 \\
        \hline
\end{tabular}

\end{table}

\begin{table}
\caption{Best fit parameters obtained by fitting the spectrum of each segment on CCD for Epoch B using Model 3: \texttt{constant*TBabs*(comptb+diskbb)}. Here, $\Gamma$: photon index, $kT_{\mathrm{e}}$: electron temperature, $kT_{\mathrm{s}}$: seed photon temperature, $N_{comp}$: normalisation of the Comptonized component, $\tau$: optical thickness, $kT_{\mathrm{in}}$: disk temperature, $N_{\mathrm{dbb}}$: disk normalization, $R_{\mathrm{in}}$: inner disk radius in km, $\dot{m}$: mass accretion rate in unit of $\mathrm{g\; cm^{-2}\; s^{-1}}$. $F_{\mathrm{bol}}$, $F_{\mathrm{tot}}$, and $F_{\mathrm{disk}}$ are the unabsorbed bolometric flux in 0.01-200 keV energy range, total unabsorbed flux and unabsorbed disk flux in 0.7-20.0 keV energy range respectively, in the unit of $\mathrm{erg \;cm^{-2} \;s^{-1}}$. $\mu$: magnetic dipole moment, B: magnetic field strength, ($R_{max}-R_{*}$): thickness of BL and $R_{max}$: radial extent of the BL.}
\label{table comptb B}
\begin{tabular}{cccccccc}
\hline
component & parameter & BB1 & BB2 & BB3 & BB4 & BB5 & BB6 \\ \hline
    comptb  & $\Gamma  = (1-\alpha)$ & $2.10_{-0.09}^{+0.11}$ & $2.02_{-0.07}^{+0.09}$ & $2.36_{-0.14}^{+0.18}$ & $2.14_{-0.12} ^{+0.16}$ & $2.30_{-0.09}^{+0.11}$ & $2.14_{-0.12}^{+0.17}$  \\
        & $kT_e$ (keV) & $3.46_{-0.13}^{+0.18}$ & $3.34_{-0.10}^{+0.13}$ & $3.74_{-0.28}^{+0.46}$ & $3.83_{-0.25}^{+0.39}$ & $3.62_{-0.16}^{+0.21}$ & $3.50_{-0.20}^{+0.32}$  \\
        & $kT_{\mathrm{s}}$ (keV) & $1.05_{-0.10}^{+0.10}$ & $0.93_{-0.11}^{+0.11}$ &  $1.08_{-0.08}^{+0.08}$ & $1.12_{-0.07}^{+0.07}$ & $1.09_{-0.06}^{+0.06}$ & $1.12_{-0.10}^{+0.10}$  \\
        & $N_\mathrm{comptb}(\times 10^{-2})$ & $3.3_{-0.1}^{+0.1}$ & $3.2_{-0.1}^{+0.1}$ &  $6.0_{-0.1}^{+0.1}$ & $6.8_{-0.1}^{+0.1}$ & $4.2_{-0.1}^{+0.1}$ & $8.4_{-0.3}^{+0.3}$  \\
         & $\tau$ & $8.5_{-0.8}^{+0.8}$ & $9.2_{-0.7}^{+0.7}$ &  $6.9_{-1.1}^{+1.1}$ & $7.8_{-1.1}^{+1.1}$ & $7.3_{-0.7}^{+0.6}$ & $8.2_{-1.2}^{+1.0}$  \\
          & $y-par$ & $2.0_{-0.3}^{+0.5}$ & $2.2_{-0.3}^{+0.4}$ &  $1.4_{-0.3}^{+0.6}$ & $1.8_{-0.4}^{+0.7}$ & $1.5_{-0.2}^{+0.4}$ & $1.8_{-0.4}^{+0.7}$ \\
    diskbb & $kT_{\mathrm{in}}$ (keV) & $0.61_{-0.03}^{+0.03}$ & $0.58_{-0.03}^{+0.03}$  &  $0.60_{-0.04}^{+0.04}$ & $0.59_{-0.02}^{+0.02}$ & $0.62_{-0.03}^{+0.03}$ & $0.62_{-0.04}^{+0.04}$ \\
        & $N_{\mathrm{dbb}}$ & $680_{-100}^{+140}$ & $800_{-150}^{+230}$ &  $880_{-170}^{+200}$ & $1020_{-150}^{+190}$ & $430_{-70}^{+90}$ & $850_{-160}^{+210}$ \\
        & $R_{\mathrm{in}}$ (km) & $24_{-2}^{+2}$ & $26_{-3}^{+4}$ &  $27_{-3}^{+3}$ & $29_{-2}^{+3}$ & $19_{-2}^{+2}$ & $27_{-3}^{+3}$  \\
        & $\dot{m} (\times 10^4 $ $  \mathrm{g\; cm^{-2}\; s^{-1}})$ & $2.95_{-0.01}^{+0.01}$ & $3.01_{-0.01}^{+0.01}$ &  $4.61_{-0.02}^{+0.02}$ & $5.02_{-0.01}^{+0.01}$ & $3.04_{-0.01}^{+0.01}$ & $6.30_{-0.02}^{+0.02}$ \\
     \hline
        & $F_{\mathrm{bol}}(\times 10^{-9} \mathrm{erg \;cm^{-2} \;s^{-1}})$ & $6.50_{-0.02}^{+0.02}$ & $6.61_{-0.02}^{+0.02}$&  $10.13_{-0.04}^{+0.04}$ & $11.05_{-0.04}^{+0.04}$ & $6.70_{-0.02}^{+0.02}$ & $13.83_{-0.05}^{+0.05}$ \\
        & $F_{\mathrm{Tot}}(\times 10^{-9} \mathrm{erg \;cm^{-2} \;s^{-1}})$ & $5.64_{-0.01}^{+0.01}$ & $5.75_{-0.02}^{+0.02}$&  $8.98_{-0.04}^{+0.04}$ & $9.79_{-0.03}^{+0.03}$ & $6.04_{-0.02}^{+0.02}$ & $12.42_{-0.04}^{+0.04}$  \\
        & $F_{\mathrm{comptb}}(\times 10^{-9} \mathrm{erg \;cm^{-2} \;s^{-1}})$ & $4.21_{-0.01}^{+0.01}$ & $4.39_{-0.02}^{+0.02}$ &  $7.18_{-0.04}^{+0.04}$ & $7.97_{-0.03}^{+0.03}$ & $5.08_{-0.01}^{+0.01}$ & $10.41_{-0.05}^{+0.05}$  \\
        & $F_{\mathrm{disk}}(\times 10^{-9} \mathrm{erg \;cm^{-2} \;s^{-1}})$  &$1.43_{-0.01}^{+0.01}$ & $1.35_{-0.01}^{+0.01}$ &  $1.80_{-0.02}^{+0.02}$ & $1.82_{-0.01}^{+0.01}$ & $0.95_{-0.01}^{+0.01}$ & $2.01_{-0.01}^{+0.01}$   \\
       
        & $F_{comptb}/F_{tot}$ & 0.75 & 0.77 &  0.80 & 0.81 & 0.85 & 0.83  \\
         & $F_{disk}/F_{tot}$ & 0.25 & 0.23 &  0.20 & 0.19 & 0.15 & 0.17   \\

          & $\mu (\times 10^{26}\mathrm{G \; cm^3})$ & 6 & 7 &  9 & 10 & 4 & 10 \\
          & B ($\times 10^{8}$G)& 15 & 18 &  24 & 27 & 10 & 27  \\
          & ($R_{max}-R_{*}$) (km) & 7 & 7 & 11 & 12 & 7 & 16  \\ 
        & $R_{max}$ (km) & 16 & 16 & 20 & 21 & 16 & 25  \\ 
          & $R_{m}$ (km) & 21 & 23 & 24 & 25 & 16 & 24  \\ 
         \hline
         
        & $\chi^2/dof$ & 87.06/104 & 99.09/104 &  80.20/90 & 76.03/99 & 90.68/98 & 65.44/99  \\
        \hline
\end{tabular}
\end{table}

\section{Results}\label{Results and discussion}
\subsection{Timing behavior of the source}
From the rms-normalized PDS, we observed VLFN, LFN, HFN, and kHz QPOs. We observed the centroid frequency of LFN  in the range of $\sim 0.9 - 7.9$ Hz. The average rms of LFN in Epoch A was found to be $\sim$ 11.1\% during Epoch A. In Epoch B, it ranged between $\sim$ 5.5\% and 10.6\%. In Epoch C, as the source moves from BC1 to BC2, the LFN component disappears, and only VLFN remains (see Table \ref{tab:Timing}).  In addition, the rms-normalized PDS showed a sharp, narrow peak at $\sim 710$ Hz ($3.7 \sigma$) and $\sim 740$ Hz ($3.7 \sigma$) in segments BB1 and BB2, respectively (see Figure \ref{PDS}).  The Q values of these two kHz QPOs are $\sim$ 12 and $\sim$ 17, respectively.  As the source moved from BB1 to BB2, the frequency of kHz QPO increased, and the fractional rms amplitude increased slightly from $\sim$ 5.7\% to $\sim$ 6.2 \%. Another signal was observed at $\sim$ 1026 Hz with a Q value of $\sim 12$ in segment BB1, which was not statistically significant ($2.4 \sigma$). In addition, to better understand the energy dependence of these kilohertz QPOs, we further generated PDS in the hard (5.0-50.0 keV) and soft (3.0-5.0 keV) energy bands. The kHz QPOs are exclusively detected in the hard energy band with an increasing rms value from $5.3_{-1.0}^{+1.0}$ \% to $7.3_{-1.2}^{+1.2}$ \% as the source transitioned from BB1 to BB2, which suggests a close relationship between the high energy emission mechanism and the origin of kHz QPOs. Furthermore, in BB6, a LF-QPO signal was observed at a frequency of 4.6 Hz with a Q-value of around 7.2. Due to the low significance (2.6 $\sigma$), we did not further use this LF-QPO in this work.

\subsection{Spectral behavior of the source}

The broad-band spectral analysis of 4U 1820-30 has been carried out in the 0.7-20.0 keV energy range using joint observation with the AstroSat SXT and LAXPC20 instruments.  The best-fit spectral parameters obtained from Model 3 are listed in Table \ref{table comptb AC} and Table \ref{table comptb B}. The best-fit spectra using Model 3 of 4U 1820-30 for segments BA1 and BB1 are shown in Figure  \ref{fig:spectra}. The parameters characterizing the thermal Comptonization component are the electron temperature ($kT_{\mathrm{e}}$), the power-law photon index ($\Gamma$), and the seed photon temperature ($kT_{\mathrm{s}}$), and they exhibit certain trends across the CCD position. Throughout the observations, it was found that the spectral state remained soft. In Epoch A, the photon index $\Gamma$ is found around $\sim$ 1.74. In Epoch C, there is marginal evidence for a decrease from $\sim 2.15$ to $\sim 1.92$ as the source moves from BC1 to BC2. During Epoch B, $\Gamma$ ranged from $\sim 2.02$ to $\sim 2.36$. The electron temperature $kT_{\mathrm{e}}$ ranged in between $\sim 2.58$ keV and $\sim 2.73$ keV in Epoch A, with an average $\sim 2.74$ keV in Epoch C. In Epoch B, when the source transitioned from BB1 to BB6, $kT_{\mathrm{e}}$ varied between  $\sim 3.34$ keV and $\sim 3.83$ keV, indicating that the electron cloud is cold during the soft state of this source. In Epoch A and Epoch C, $kT_{\mathrm{s}}$ is roughly constant at 0.90 keV and 0.94 keV, respectively; however, in Epoch B, it varies between $\sim 0.93$ and $\sim 1.12$. On the other hand, we have estimated the inner disk temperature ($kT_{\mathrm{in}}$) from Model 3. The inner disk temperature remains approximately constant at $\sim 0.6$ keV across the segments within uncertainties.  \\

\subsection{Estimation of physical parameters}
 We have derived certain key physical parameters from Model 3, such as the inner accretion disk radius ($R_{\mathrm{in}}$), optical depth ($\tau$), and the Comptonization parameter ($y_{par}$). To calculate all physical parameters, we have assumed the source distance ($D_{10}; \text{in unit of 10 kpc})$to be 8 kpc \citep{Baumgardt2021,marino2023accretion, IXPE2023}, disk inclination angle ($\theta)=40^\circ$ \citep{anderson1997time}, mass of the NS ($M_{\mathrm{NS}})=1.58 M_\odot$ \citep{guver2010mass} and the radius of the NS ($R_{\mathrm{NS}}$)=9.11 km \citep{guver2010mass}. The normalization of the multi-color disk blackbody ($N_{\mathrm{dbb}}$) component of the disk gives an estimate of $R_{\mathrm{in}}$ \citep{Mistuda,Makishima}. It is given by,  
\begin{equation}
    \label{radius}
    N_{\mathrm{dbb}} = \left(\frac{R_{in}}{D_{10}}\right)^2 cos\theta 
\end{equation}
We find that $R_{\mathrm{in}}$ lies approximately in the range of $\sim$ 19-40 km, indicating that the accretion disk is truncated. In Epoch A, it was found to vary between $28_{-4}^{+6}$ and $40_{-4}^{+8}$ km, while in Epoch C, it was found to be $\sim 33$ km. During Epoch B, $R_{\mathrm{in}}$ ranged between $19_{-2}^{+2}$ km and $29_{-2}^{+3}$ km. Assuming spherical geometry and a uniform density of the corona, we derived the optical depth ($\tau$) from the expression of spectral index ($\alpha$) \citep{Zdziarski1996} as given by,
\begin{equation}
    \label{opticadepth}
 {\alpha}= \left[\frac{9}{4}+\frac{m_ec^2}{kT_{e}\times \tau \left(1+\frac{\tau}{3}\right)}\right]^{1/2} - \frac{3}{2}
\end{equation}
 where $m_{e}$ is the mass of an electron, c is the speed of light in a vacuum, and $\tau$ is the optical depth. The corona was found to be optically thick in all segments. The average optical depth of the corona was found to be approximately 13 in Epoch A and 10.2 in Epoch C, while it ranged between $\sim 7$ and $\sim 9.2$ in Epoch B. The Comptonization parameter ($y_{par}$), which measures the degree of Comptonization, is calculated using Equation \ref{y-para} \citep{Agarwa2020},
\begin{equation}
\label{y-para}
  y_{par} = \frac{4kT_{e}}{m_{e}c^2} \tau^2  
\end{equation}
In Epoch A and Epoch C, the $y_{par}$ was roughly constant at 3.5 and 2.2, respectively. However, it varied from $\sim 1.4$ to $\sim 2.2$ in Epoch B.

In addition, we have estimated the total unabsorbed bolometric flux ($F_{\mathrm{bol}}$) in the energy range 0.01-200 keV, the unabsorbed total (${F_{total}}$), unabsorbed disk ($F_{\mathrm{disk}}$) and unabsorbed \texttt{comptb} ($F_{comptb}$) fluxes in the energy range 0.7-20.0 keV using a multiplicative model \texttt{cflux} in Model 3. We found the Comptonization component contributes around 80\% of the total flux. To understand the possible reason for the disk truncation, we calculate certain physical parameters, including the magnetic dipole moment ($\mu$), the magnetic field strength (B) at the NS poles, the radius of the BL ($R_{max}$) and the thickness of the BL $(R_{max}-R_{*})$. Assuming the obtained truncated accretion disk is due to the magnetic field of the NS system \citep{illarionov1975number}, the upper limit of the magnetic dipole moment $\mu$ $(\mathrm{\,G \,cm^3})$ was calculated following \cite{cackett2009broad}:
\begin{equation}
    \label{magnetic}
\mu = 3.5 \times 10^{23} \times \left(\frac{\mathrm{x}}{k_A}\right)^{7/4} \times \left(\frac{M_{NS}}{1.4 M_\odot}\right)^2 \times \left(\frac{f_{ang}}{\eta} \frac{F_{bol}}{10^{-9} \mathrm{\;erg\; cm^{-2} \;s^{-1}}}\right)^{1/2} \times D_{3.5} \mathrm{\;G\;cm^{3}}
\end{equation}
where, $k_A$ is a factor influenced by the transition from spherical to disk accretion, $f_{ang}$ represents the anisotropy correction factor, $\eta$ denotes the Schwarzschild metric's accretion efficiency, and $\mathrm{x}$ is defined by the relation $R_{in}=\mathrm{x}GM_{NS}/c^2$ \citep{cackett2009broad}. The magnetic field strength (B) at the poles, B, was calculated by dividing the magnetic dipole moment by $R_{NS}^3$ \citep{ludlam2017truncation}, assuming $k_A=1$, $f_{ang}=1$, and $\eta=0.1$ for a non-pulsating NS. As shown in Table \ref{table comptb AC} both $\mu$ and B at the poles of the NS 4U 1820-30 decreased as the source moved across the BS in Epoch A and Epoch C. In Epoch A, $\mu$ decreased from $\sim$ 29 to 11 $(\times 10^{26}\mathrm{\,G \,cm^3})$, while in Epoch C it varied from $\sim$ 15 to 12 $(\times 10^{26}\mathrm{\,G \,cm^3})$. In Epoch B, $\mu$ varied between $\sim$ 4 and 10 $(\times 10^{26}\mathrm{\,G \,cm^3})$, also as reported in Table \ref{table comptb B}. 

\begin{figure} [h!]
    \centering
    \gridline{\fig{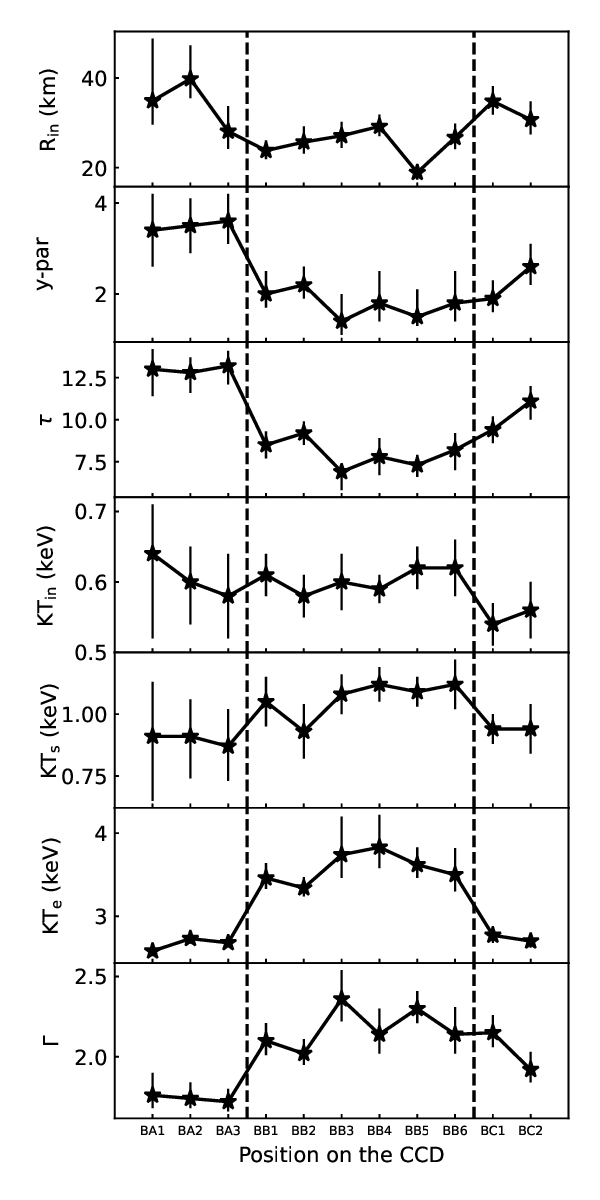}{0.4\textwidth}{}
              \fig{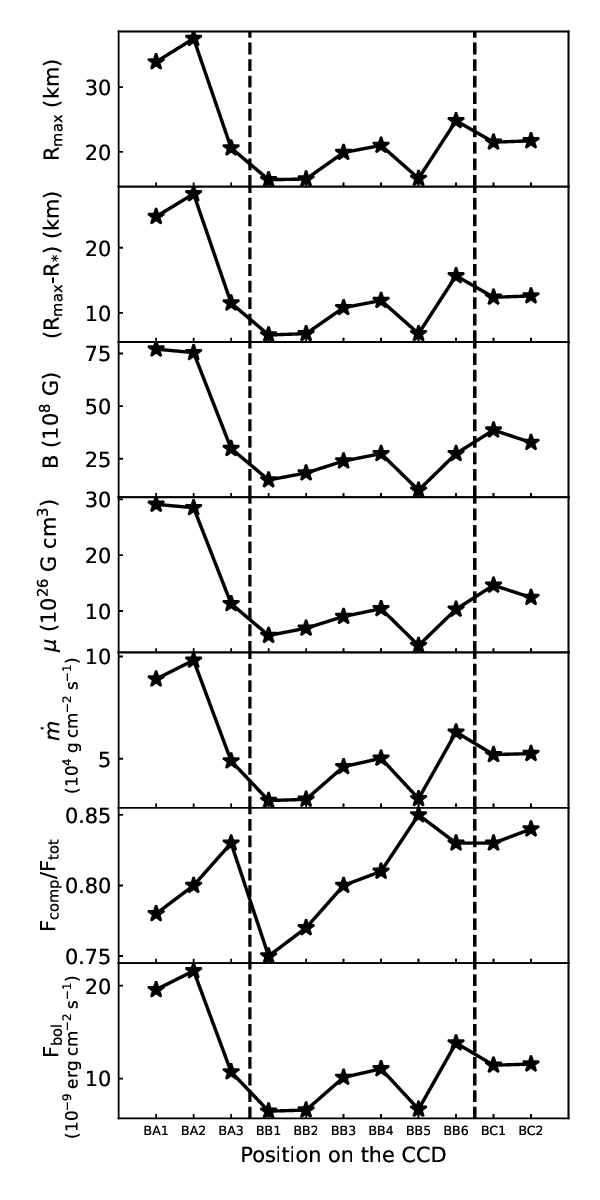}{0.4\textwidth}{}}    
    \caption{Evolution of the spectral parameters for 4U 1820-30 along the CCD, obtained using Model 3: \texttt{constant*tbabs*(Comptb+diskbb)} applied to the combined SXT and LAXPC20 data. From left to right, the vertical dashed lines indicate the transitions between Epoch A and Epoch B, and between Epoch B and Epoch C.}
    \label{ALL correlations}
\end{figure}

Another possible mechanism for disk truncation is the presence of BL. In a geometrically thin accretion disk, matter rotates near the Keplerian velocity in the innermost region. When the Keplerian velocity exceeds the rotational velocity of the NS, a BL forms where the excess kinetic energy is dissipated \citep{Syunyaev,inogamov1999spread}. This region is hotter than the surrounding accretion disk, producing more energetic radiation from the compact region near the NS, and can cause Comptonization of seed photons \citep{revnivtsev2006boundary}. The maximum radial extent of the BL ($R_{max}$) was estimated following \cite{popham2001accretion}:
 \begin{equation}
 \label{BL}
     \log (R_{max}-R_{*}) \simeq 5.02+0.245 \left[\log \left(\frac{\dot{m}}{10^{-9.85} M_{\odot} y^{-1}}\right)\right]^{2.19}
 \end{equation}
where ($R_{max}-R_{*}$) and $R_{*}$ are the thickness of BL and radius of the, NS, respectively. 
To estimate the $R_{max}$, $\dot{m}$ ($\mathrm{\;g \;cm^{-2} \;s^{-1}}$) at the surface of the NS was first calculated following \cite{Galloway2008}:
\begin{equation}
 \label{massrate}
   \dot{m} = 6.7\times10^3\times \frac{F_{bol}}{10^{-9} \mathrm{erg\; cm^{-2}\; s^{-1}}} \times \left(\frac{d}{10 \;\mathrm{kpc}}\right)^2  
       \times \left(\frac{M_{NS}}{1.4 \; M_\odot}\right)^{-1}  \times \left(\frac{1+\mathrm{z}}{1.31}\right) \times \left(\frac{R_{NS}}{10 \; \mathrm{km}}\right)^{-1}  \mathrm{g\; cm^{-2}\; s^{-1}}
 \end{equation}
where, $\mathrm{z}$ represents the surface redshift, determined as $1+\mathrm{z}=\left(1-\frac{2GM_{NS}}{R_{NS}c^2}\right)^{-1/2}$, $G$ and $c$ are the universal gravitational constant and speed of light in the CGS system, respectively.  We have calculated the surface redshift of the NS 4U 1820-30 to be $\sim$ 0.43. The value of $\dot{m}$ was found to vary between  $4.89_{-0.01}^{+0.01} \times 10^4 $ $  \mathrm{g\; cm^{-2}\; s^{-1}}$ and $9.82_{-0.03}^{+0.03} \times 10^4 $ $  \mathrm{g\; cm^{-2}\; s^{-1}}$ in Epoch A. In Epoch C, it remains approximately constant around $5.2 \times 10^4 $ $  \mathrm{g\; cm^{-2}\; s^{-1}}$, while in Epoch B it varies between $2.95_{-0.01}^{+0.01} \times 10^4 $ $  \mathrm{g\; cm^{-2}\; s^{-1}}$ and $6.30_{-0.02}^{+0.02} \times 10^4 $ $  \mathrm{g\; cm^{-2}\; s^{-1}}$. Based on the inferred value of $\dot{m}$, the obtained values of $R_{max}$ and ($R_{max}-R_*$) are listed in Table \ref{table comptb AC} for Epoch A and C, and in Table \ref{table comptb B} for Epoch B. To understand the spectral evolution of the source along the CCD, we plot spectral parameters $\Gamma$, $kT_{e}$, $kT_{s}$, $\tau$, $kT_{in}$, $y-par$, $R_{in}$, $F_{bol}$, $F_{comp}/F_{tot}$, $\dot{m}$, $B$, the boundary layer thickness $R_{max}-R_{*}$, and $R_{max}$ as a function of the source position on the CCD. The variations of these parameters are shown in Figure \ref{ALL correlations}.

\section{discussion} \label{discusion}
In this work, we performed a detailed long-term spectral and temporal analysis of the atoll source 4U 1820-30 using the SXT and LAXPC-20 instruments on-board \textit{AstroSat}. The HID showed that all AstroSat observations of this source are part of BS, as shown in Figure \ref{maxi_bat_hardness}. To conduct the spectral and temporal analysis, we have divided the complete set of data into three different Epochs.
During all three Epochs, the source is found in the banana state. Subsequently, we divided the data set into 11 segments based on the source's position on the CCD (see right panel of Figure \ref{fig:lightcurve and CCD}). Our complete observations did not reveal any type-I thermonuclear X-ray bursts. From the fast timing analysis, the PDS of each segment is modeled by two components: power law and Lorentzian in the 3.0-50.0 keV energy range using LAXPC 20 data. This PDS are mainly composed of LFN, HFN, and VLFN. In addition to this, the timing analysis revealed the presence of the kHz QPO (see Figure \ref{PDS}) and its evolution from $\sim 710$ Hz (in BB1) to $\sim 740$ Hz (in BB2). Broadband spectral data acquired with SXT and LAXPC 20 were fitted with the two widely accepted approaches in the 0.7-20.0 keV energy range for each segment. The details of spectral analysis are described in section \ref{spectral_analysis}. Among these approaches, the spectral data were best described by a combination of the MCD \citep{Mistuda} and a thermal Comptonization model \citep{comptb}. The MCD model explains the emission from the accretion disk, while the thermal Comptonization model accounts for the emission from the corona/BL. The variation of the spectral parameters and their association with the kHz QPO properties are also studied.

Based on analysis of the broadband spectra, we found that $\Gamma$ spans $\sim$1.72 to 2.38, indicating that 4U 1820-30 was likely in a BS or soft state. When the source was in the soft state, \cite{Titarchuk}, \cite{Mondal:2016yez}, and \cite{marino2023accretion} found a very close $\Gamma$ value, consistent with our outcomes. Conversely, \cite{agrawal20184U170544} found that $\Gamma < 2.0$, when the source 4U 1705-44 was in UB. We have found that the electron temperature of the corona varies between $\sim$ 2.58 and 3.83 keV, which is lower than the value when the source was in IS \citep{Titarchuk}. During BS, the value of the electron temperature of this source is consistent with the previously reported results (see, \cite{Titarchuk,Mondal:2016yez,IXPE2023,marino2023accretion}).  To better understand the coronal properties, we have estimated the optical thickness of the corona, which has been found to vary from $\sim$ 7 to 13 and aligns well with earlier studies \citep{kaaret1997strong,piraino1999bepposax}. The optical thickness of the corona depicted an increasing trend as the source moved in the BS during three distinct Epochs (see Figure \ref{ALL correlations}). Ultimately, it was determined that the corona is cool and optically dense by presuming the seed photon to originate from blackbody radiation at the NS surface. The values $kT_e$ and $\tau$ are also the same in other atoll-type sources during BS, such as 4U 1608-52 \citep{sree} and 4U 1705-44 \citep{agrawal20184U170544}.

We found that the inner accretion disk was truncated and ranged from $\sim$ 19 km to 40 km. The inner disk temperature varied between $\sim$ 0.54 keV to 0.64 keV, which is very close to the previously reported results (see \cite{marino2023accretion,IXPE2023, Mondal:2016yez}). This truncation of the inner rim of the accretion disk may be caused by the magnetic field strength \citep{illarionov1975number}. In 4U 1705-44, \cite{agrawal20184U170544} suggested that an increase in optical depth along the banana branch is caused as radiation pressure drives material from the accretion disk onto an extended corona. This interpretation aligns well with our findings. On the other hand, \cite{Mondal:2016yez, Russell2021, marino2023accretion, IXPE2023} suggested the presence of BL between the inner rim of the accretion disk and the surface of the NS. \cite{agrawal20184U170544} discussed that as the accretion disk radius approaches the NS surface, the BL/central corona becomes optically thick. So, we assumed the presence of an optically thick BL or a cold central corona present in the space between the NS surface and the inner rim of the accretion disk. For further understanding, we have estimated the thickness and the radial extent of the BL. The thickness of the BL ranged between $\sim 7$ km to $\sim 28$ km and the radial extent of the BL ranged between $\sim 16$ km and $\sim 38$ km. In addition, we have also observed that nearly 80\% of the total flux was dominated by the BL/central corona. As shown in Figure \ref{ALL correlations}, the variation in physical parameters along the CCD's position reveals that, in Epoch A, as the $R_{in}$ drops, the optical thickness and electron temperature remain constant. The corona is found to be optically thick ($\tau \sim 13$) and cool ($kT_{e} \sim 2.7$ keV).  This suggests that the photons undergo a large number of scatterings and hence it become cooler.  On the other hand, as $R_{\mathrm{in}}$ decreases, the thickness of the cold central corona/BL also decreases, but the radial extent of the BL is very close to the $R_{\mathrm{in}}$. Similarly, in Epoch C, as $R_{in}$ slightly decreases, the optical thickness of the central corona was found to be around 10.2 with an electron temperature of around 2.74 keV. Our analysis reveals that the BL extends radially close to the accretion disk's inner edge, except BB1 and BB2 segments. This may be attributed to variations in magnetic strength, which leads to disk truncation and results in the BL or central corona cooling down. To understand the cause behind the disc truncation due to the magnetic field associated with the NS, we further calculated the upper limit Alfven radius ($R_{A}\equiv(\frac{\mu^4}{2GM\dot{M}^2})^{1/7}$)  and magnetospheric radius ($R_{m}=\xi R_{A}$) assuming $\xi=0.6$ \citep{li1997evolution}. $R_{A}$ has been calculated by using the obtained $\mu$ value listed in Table \ref{table comptb AC} and Table \ref{table comptb B}. We found that $R_{in}$ is always slightly greater than or equal to $R_{m}$, which means that the disk is truncated by the magnetic field. The $R_{max}$ is always less than $R_{m}$, which means that BL lies within the $R_{m}$. Interestingly, when the source moves from the BB1 to the BB2 segment, the $kT_{\mathrm{e}}$  around $3.4$ keV, $kT_{s}$ around $ 1.0$ keV. The $R_{\mathrm{in}}$ and $kT_{\mathrm{in}}$ also show no significant variation. However, the optical thickness of the central corona/BL increases from $\sim 8.5-9.2$. Furthermore, in these segments, the location of the inner rim of the accretion disk and $R_{max}$ do not coincide, and the $R_{\mathrm{in}}$ is slightly higher than the size of the BL/central corona (see Table \ref{table comptb B} and Figure \ref{ALL correlations}). This suggests that there is no short transition between the disk and BL. The region between the BL and the inner disk is filled with sub-Keplerian flow,  which suggests that the accretion flow geometry is different during segments BB1 and BB2.

In addition, we identified kHz QPOs in these segments where the mass accretion rate was significantly low. From the previous finding, in other atoll-type sources the kHz QPOs was observed at the lowest mass accretion rate in the BS, which aligns with our result \citep{wijnands1999broadband,psaltis1999correlations,di2001study, Salvo2003}. The frequency from $\sim 710$ Hz (in BB1) to $\sim 740$ Hz (in BB2) indicates the evolution of the kHz QPO at the lowest mass accretion rate around $2.9\times10^4$ $  \mathrm{g\; cm^{-2}\; s^{-1}}$. Besides, as noted by \cite{wang}, kHz QPOs in most atoll-type sources typically occur in the LLB branch, and similarly, this pattern is observed in our data (refer to the middle panel of the CCD in Figure \ref{fig:lightcurve and CCD}). Previously, 4U 1820-30 showed twin kHz QPOs in the BS branch \citep{Smale1997,zhang,Titarchuk}. On the other hand,  in the BB1 segment, another signal around $\sim 1025.6$ Hz was also observed, which was statistically insignificant. Additionally, we have noticed that these lower kHz QPOs are identified within the 5-50 keV energy range, implying a likely origin from the high-energy Comptonized component. To investigate this, several models have been proposed to explain this kind of oscillation \citep{Alpar1985, Lamb1985, Titarchuk1998, Stella1998, Stella1999, Wijnands2003}. One of the prominent relativistic precession models (RPM) is used to understand the behaviour of lower kHz QPO frequency. Using the RPM model, we have calculated the radius where the kHz QPOs originate. To calculate this, we have used the following Equation of \cite{Stella1998},\\
\begin{equation}
\label{RPM}
\begin{split}
  \nu_\phi &=\pm M_\mathrm{NS}^{1/2}r^{-3/2}[2\pi(1\pm aM_\mathrm{NS}^{1/2}r^{-3/2})]^{-1} \\
  \nu_r^2 &= \nu_\phi^2 (1-6M_\mathrm{NS}r^{-1}\pm 8aM_\mathrm{NS}^{1/2}r^{-3/2}-3a^2r^{-2})
\end{split}
\end{equation}

where $\nu_\phi$ and $\nu_r$ are the azimuthal and epicyclic frequencies (G=c=1). $M_{NS}$ and $a$ correspond to the mass and angular momentum of the NS, and $r$ is the radial distance. In the RPM model, the lower frequency kHz QPOs are identified with the periastron precession frequency ($\nu_{per}=\nu_\phi - \nu_r$). We have assumed the `Schwarzschild limit' ($a = 0$) and estimated that the frequencies of $\sim 710$ Hz and $\sim 740$ Hz originate at a radial distance of $\sim 16$ km from the NS surface. From the spectral analysis, the radial extent of the BL for BB1 and BB2 is also found to be $\sim$ 16 km; both results are nearly identical. Earlier, \cite{wang2013} calculated the Keplerian orbital radii to be approximately 15.6 km based on the upper kHz QPO measurement for this source using the RPM model. \cite{Stella1998,Stella1999} suggested that kHz QPOs originate at the transition boundary between the optically thick disk and the hot inner region. They proposed that these could be due to occultation by orbiting blobs, and \cite{vietri1998new} suggested the resonant z-oscillations of the blobs. They also suggested that if a magnetosphere is present, it should only minimally affect the motion of the blobs. Our findings indicate that both kHz QPOs in BB1 and BB2 segments emanate from the region between the surface of the NS and the inner rim of the accretion disk. Therefore, we propose that the area between the accretion disk and the NS surface may contain an optically thick boundary layer or central corona, whose oscillations could be responsible for generating these kHz QPOs.

\section{Conclusion}
In this section, we present the main findings derived from the X-ray analysis of 4U 1820-30 using the combined data from \textit{AstroSat} instruments SXT and LAXPC 20.
\begin{itemize}
    \item A long-term comprehensive study of atoll type NSXRB 4U 1820-30 using \textit{AstroSat}'s observations was performed. The source was found in the banana state during these observations. 
    \item The spectral parameters generally align with what has been documented in prior studies. The spectra primarily involve two components: one originating from the accretion disk and the other from the Comptonization mechanism. The spectrum is predominantly influenced by the Comptonized component.  As the inner edge of the accretion disk nears the surface of the NS, the optical thickness of the corona/boundary layer increases. The primary cause for this truncation of the disk’s inner rim is the magnetic field strength.
    \item Detection of kHz QPO in BB1 and BB2 segments presents fascinating insights. Particularly in the context of the evolution of kHz QPOs. The radial extent of the boundary layer from spectral observations closely matches the radius predicted by the RPM model. This finding implies that the region between the inner edge of the accretion disk and the neutron star's surface might harbour an optically thick BL or central corona, whose oscillations can produce these kHz QPOs.
\end{itemize}

\section*{Acknowledgements}
\begin{acknowledgments}
We thank the anonymous referee for the useful comments and suggestions that have improved the quality of the paper enormously. PT acknowledges the financial support of ISRO under the AstroSat archival Data utilization program (No: DS-2B-13013(2)/8/2019-Sec.2). This publication uses data from the AstroSat mission of the Indian Space Research Organisation(ISRO), archived at the Indian Space Science Data Centre (ISSDC). This work has used data from the SXT and LAXPC instruments on board AstroSat. The LAXPC data were processed by the Payload Operation Center (POC) at TIFR, Mumbai. This work has been performed utilizing the calibration databases and auxiliary analysis tools developed, maintained, and distributed by the AstroSat-SXT team with members from various institutions in India and abroad and
the SXT POC at the TIFR, Mumbai (\url{https://www.tifr.res.in/~astrosat_sxt/index.html}). The SXT data were processed and verified by the SXT POC. This research has also made use of data supplied by the UK Swift Science Data Centre at the University of Leicester and MAXI data provided by RIKEN, JAXA, and the MAXI team. S.D. acknowledges Guru Ghasidas Vishwavidyalaya (Central University), Bilaspur, Chhattisgarh, India, for providing a fellowship under the VRET scheme. S.D. also expresses his sincere thanks to Mr. Sayan Bandyopadhyay for his invaluable help and support. P.T. expresses his sincere thanks to the Inter-University Centre for Astronomy and Astrophysics (IUCAA), Pune, India, for granting support through the IUCAA associateship program. V.K.A. thanks GH, SAG; DD, PDMSA, and Director, URSC for encouragement and continuous support to carry out this research.
\end{acknowledgments}

%

\vspace{5mm}
\section*{Data Availability}

Data used for this publication are currently available at the Astrobrowse (AstroSat archive) website (\url{https://astrobrowse.issdc.gov.in/astro_archive/archive/}) of the ISSDC.\\

\textit{Facilty:} AstroSat (LAXPC and SXT)
\textit{Software:} LaxpcSoft, HEASoft (V. 6.31), XSPEC (12.13.0c; \citep{arnaud1996astronomical}) , GNUPLOT, matplotlib \citep{Hunter2007}.




\bibliography{PT_references}{}
\bibliographystyle{aasjournal}




\end{document}